\documentclass[lettersize,journal]{IEEEtran}
\usepackage{amsmath,amsfonts}
\usepackage{algorithmic}
\usepackage{algorithm}
\usepackage{array}
\usepackage[caption=false,font=normalsize,labelfont=sf,textfont=sf]{subfig}
\usepackage{textcomp}
\usepackage{stfloats}
\usepackage{url}
\usepackage{verbatim}
\usepackage{graphicx}
\usepackage{cite}
\usepackage{xcolor}
\usepackage{quotes}
\usepackage{placeins}

\usepackage[normalem]{ulem} 
\usepackage{graphicx} 
\usepackage{epstopdf} 
\usepackage{tikz-cd}
\DeclareGraphicsExtensions{.eps} 
\usepackage{microtype} 
\usepackage{eso-pic} 

\begin{document}

\AddToShipoutPictureFG*{%
  \AtPageUpperLeft{%
    \raisebox{-1.1cm}{
      \makebox[\paperwidth]{%
        \begin{minipage}{0.92\paperwidth}
          \centering
          \color{gray}

          {\fontsize{7}{8}\selectfont
          This is the authors' final version of the manuscript published as:
          Spisak, J., Popescu, S. T., Wermter, S., Hoffmann, M. and O'Regan, J. K. (2026),\\
          'A computational model of infant sensorimotor exploration in the mobile paradigm',
          IEEE Transactions on Cognitive and Developmental Systems 18 (3), 692-707.
          \textcopyright~IEEE,\\
          \texttt{https://doi.org/10.1109/TCDS.2025.3635281}\\
          \par}
        \end{minipage}%
      }%
    }%
  }%
}

\title{A computational model of infant sensorimotor exploration in the mobile paradigm}

\author{Josua Spisak$^{1,\dagger}$, Sergiu Tcaci Popescu$^{2,\dagger,*}$, Stefan Wermter$^{1}$, Matej Hoffmann$^{2,*}$, J. Kevin O'Regan$^{3}$

\thanks{
S.T.P. and M.H. were supported by the Czech Science Foundation (GA CR), project no. 25-18113S. S.T.P. was additionally supported by the project Mobility \v{C}VUT MSCA-F-CZ-I, no. CZ.02.01.01/00/22\_010/0003405. J.S. and S.W. were supported by the DFG (MoReSpace), part of the DFG Priority Program 2134 ``The Active Self’’ (DFG PA2302/13-1).}
\thanks{$^{1}$Josua Spisak and Stefan Wermter are with the University of Hamburg, Informatics Department, Knowledge Technology Group, Germany.} 
\thanks{$^{2}$Sergiu T. Popescu and Matej Hoffmann are with Department of Cybernetics, Faculty of Electrical Engineering, Czech Technical University in Prague, Czech Republic.} 
\thanks{$^{3}$J. Kevin O'Regan is with Université Paris Cité and CNRS, Paris, France.}
\thanks{$^\dagger$ Equal contribution.}
\thanks{$\textbf{*}$ Corresponding authors:
Matej Hoffmann {\tt\small matej.hoffmann@fel.cvut.cz} and Sergiu T. Popescu
{\tt\small sergiu.t.popescu@gmail.com}.}
}


\maketitle
\begin{abstract}
We present a computational model of the mechanisms that may determine infant behavior in the ``mobile paradigm''. This paradigm has been used in developmental psychology to explore how infants learn the sensory effects of their actions. In this paradigm, a mobile (an articulated and movable object hanging above an infant's crib) is connected to one of the infant's limbs, prompting the infant to preferentially move that ``connected'' limb. This ability to detect a ``sensorimotor contingency'' is considered to be a foundational cognitive ability in development. To understand how infants learn sensorimotor contingencies, we built a model that attempts to replicate infant behavior. Our model incorporates a neural network, action-outcome prediction, exploration, motor noise, preferred activity level, and biologically inspired motor control. We find that simulations with our model replicate the classic findings in the literature showing preferential movement of the connected limb. An interesting observation is that the model sometimes exhibits a burst of movement after the mobile is disconnected, shedding light on a similar occasional finding in infants. In addition to these general findings, the simulations also replicate data from two recent more detailed studies using a connection with the mobile that was either gradual or all-or-none. A series of ablation studies further shows that the inclusion of mechanisms of action-outcome prediction, exploration, motor noise, and biologically inspired motor control was essential for the model to correctly replicate infant behavior. This suggests that these components are also involved in infant sensorimotor learning.\footnote{The code of the model is published at: \url{https://github.com/ctu-vras/mobile-paradigm-model.}}
\end{abstract}


\section{Introduction}
For a developing infant, detecting the effect of its own actions on its sensory input (i.e. detecting ``sensorimotor contingencies'') is presumably a core mechanism by which the infant first learns how to distinguish its body from its environment, and then learns how to control the body and use it to grasp and manipulate external objects (for reviews, see \cite{jacquey_fagard_2020} and \cite{sen_making_2021}). Becoming a social being, perceiving itself and its caregivers as agents with causal effects is also an essential aspect of infant development that requires noting contingencies between its actions and other people's reactions \cite{tarabulsy_contingency_1996},\cite{adamson_stillface_2003}.

In the literature on infant development, a variety of paradigms have been used to study detection of sensorimotor contingencies, using different types of actions on the part of the infant (limb movements, head movements, facial expressions, vocalizations) and different types of sensory consequences (sounds, voices, visual stimulation, including human or schematic faces).

In what follows, we present the key findings from mobile paradigm studies focused on sensorimotor contingency detection and learning. We then provide a brief overview of previous models that address target infant behavior and highlight the main advances and novel aspects of our model.

\subsection{The mobile paradigm}
One of the most frequently used methods to study the detection of sensorimotor contingencies has been the ``mobile paradigm'' introduced in the late 1960’s by Rovee \& Rovee \cite{rovee_conjugate_1969} and by Watson \& Ramey\cite{watson_reactions_1972}. In the classic version, a ``mobile'' (a movable, articulated object with attached decorations) that hangs above an infant's crib is connected by a ribbon to one of the infant’s legs or arms so that the movement of the connected limb influences the motion of the mobile. 

Early studies demonstrated that infants from at least 2 months onward can rapidly detect the contingency between their actions and the mobile (e.g. \cite{greco_ontogeny_1986},\cite{hayne_ontogeny_1986}, \cite{lewis_emotional_1985}, \cite{alessandri_violation_1990}), showing more motor activity than when there is no connection. Five- to six-month-old infants start to reliably distinguish which is the ``connected'' limb and move it more than the other limbs \cite{jacquey_popescu_2020}, \cite{popescu_6-month-old_2021}. Later research by Rovee-Collier's group (e.g. \cite{rovee-collier_development_1999},\cite{rovee-collier_long-term_1999}) turned away from contingency detection itself. Instead, they used the phenomenon to test the duration of retention of the contingency, the particular visual cues that an infant will use, how the infant will generalize to other cues and environmental parameters.

Recently several studies have gone back to more carefully reviewing the literature and re-investigating the basic contingency detection phenomenon itself to confirm the exact conditions and time-course of the infant's reaction to the establishment of a contingency \cite{sen_making_2021}, \cite{jacquey_popescu_2020}, \cite{popescu_6-month-old_2021},\cite{sen_methodological_2024}, \cite{sloan_meaning_2023}. This recent work has shown that infants' ability to detect contingencies is not as easy to demonstrate as the earlier literature might have led one to believe. However, the main empirical results that appear to be reliable are the following.  

\textbf{Specificity of connected limb:} For infants aged 5-6 months, when one limb is connected to the mobile, that limb's activity rapidly becomes overall higher than that of the other limbs. This effect appears within the first minute of the experiment, showing that infants detect the contingency very quickly and are able to activate the appropriate limb selectively \cite{watanabe_general_2006}, \cite{popescu_6-month-old_2021}.

\textbf{Comparison with a control group:} In addition to the main experimental group where the mobile is activated contingently on the infant's movements, some studies use a control group where the mobile is activated non-contingently, i.e. with a similar activation rate but in a random fashion independent of the infant's movements (for example, \cite{rovee_conjugate_1969}, \cite{rovee-collier_topographical_1978}, \cite{jacquey_popescu_2020},\cite{popescu_6-month-old_2021},\cite{angulo-kinzler_three-month-old_2002}). By showing that infants' activity in the contingent group is still higher than in the non-contingent control group, these studies show that the infants are sensitive to the contingency itself, and not just the activity of the mobile.

In addition to these two main empirical results, there are two interesting but less reliable findings, which we will consider in this paper.

\textbf{Extinction burst:} When, at the end of the experiment, the connected limb is disconnected from the mobile, there sometimes is an ``extinction burst''---the activity of the connected limb briefly increases and then decreases again, to a level similar to that of the other limbs. The existence of an extinction burst is not systematically observed in the literature \cite{alessandri_violation_1990},\cite{heathcock_relative_2005},\cite{lewis_violation_1990},\cite{rovee-collier_topographical_1978}, and might be indicative of the presence of a predictive mechanism at work in the infant's behavior. 

\textbf{Greater effect for binary vs. non-binary:} When the stimulation caused by the infant's limb movements is non-binary or ``conjugate'' with the movement of the mobile---that is, when the connected limb's activity determines the mobile movement in a proportionate way---the effects described above are less clearly visible than in a ``binary'' condition, where the connected limb always triggers the maximum mobile movement when the connected limb movement passes a threshold. This finding from empirical work \cite{popescu_6-month-old_2021} is worth noting because it might be predicted from the computational model to be presented here. 
 
\subsection{Computational models of the mobile paradigm}
Surprisingly, despite the substantial literature and the importance of contingency detection as a mechanism underlying infant development, there have been few attempts to model the findings observed in the mobile paradigm, either mathematically or by computer simulation. 

An exception was the work of John S. Watson and his collaborators Butko \& Movellan~\cite{butko_detecting_2010},\cite{movellan_Watson_development_2002},\cite{butko_learningToLearn_2007}. They were struck by the fact that infants in the mobile paradigm often coo and smile in the presence of the connected mobile, suggesting that they might be putting to work a mechanism designed to seek out social contingencies. They proposed that infants in the mobile paradigm might be deploying a similar kind of action/waiting-for-reaction behavior that might be used to detect whether one is communicating with a social partner rather than a random or a completely deterministic entity (like an echo). They modeled an agent that maximizes the information gain or reduces the uncertainty and combined it with reinforcement learning to conceive an optimal infomax controller for detecting social contingencies. They showed that this approach accounts for infants' behavior in a situation in which a simple robot was animated when the infant generated vocalizations. However, as pointed out by the authors, the infomax calculation is probably not practical for a brain to make, and a simpler approximation would be more realistic. 

Two other attempts to explicitly model behavior in the mobile paradigm have also appeared recently. 

Zaadnoordijk et al.'s~\cite{zaadnoordijk_can_2018} goal was to demonstrate that the most basic effects observed in the mobile paradigm could be explained \textit{without} assuming that the infant was explicitly searching for contingencies. They were, therefore, taking the opposite viewpoint from Butko \& Movellan, trying to show that the mobile paradigm is \textit{not} a good test for a notion of causality or agency in infants. Zaadnoordijk et al.~\cite{zaadnoordijk_can_2018} set out to explain the observed infant behavior using a simple ``babybot'' model in the form of a state machine with four limbs that could each be in three different positions (top, middle, and bottom) and three possible actions (move up, move down, hold still). They showed how a very simple operant conditioning mechanism can fairly accurately simulate the temporal course of infant learning in the mobile paradigm. Their mechanism worked simply by repeating movements that produce a sensory effect, i.e. by moving the connected limb. The controller (``brain'') part of the model was deliberately designed in a minimalist and non-representational fashion---the agent had no internal machinery to represent or predict the consequences of its actions (action-effect relationship). Simple reinforcement of actions triggering the movement of the mobile was enough to reproduce the core behavior, i.e. more movement in the limb connected to the mobile. However, because their model contains no predictive element, it does not seem compatible with the possibility of an ``extinction burst'' following the removal of the contingency~\cite{rovee-collier_topographical_1978}. 

Kelso \& Fuchs~\cite{kelso_coordination_2016} also proposed a model without agency and with no predictive component. Unlike the discrete steps in the state machine in \cite{zaadnoordijk_can_2018}, Kelso \& Fuchs's approach was rooted in the dynamical systems framework (see also \cite{fujihira2023dynamical} for another model), paying close attention to the temporal dynamics and the properties of the physical movements of the infants' limbs, the mobile, and their coupling (the connected limb being physically attached to the mobile with a ribbon). The changes in the infant behavior were explained by phase transitions in this coupled dynamical system---a mechanistic explanation that does not require any rewards, reinforcements, or cognitive mechanisms (contingency detection, prediction, etc.). Since, again, no predictive element is present in this approach, it is unclear whether an ``extinction burst'' could be explained in this framework. 

\subsection{Our model}
The purpose of the present paper is to propose a first attempt at a detailed computational model, shown in Fig. \ref{fig:Flowchart}, of the mobile paradigm that addresses the main two empirical findings observed in infants at age 6 months: the increase in activity specific to the ``connected'' limb, and a significant effect for a contingent group but not for a control group with random stimulation. Additionally, we examine how the model might simulate the controversial phenomenon of the extinction burst and the difference between binary (non-conjugate) and non-binary (conjugate) stimulation. 

In our model, there were two important guiding principles that we wanted to adhere to. The first guiding principle was the fact that contrary to Zaadnoordijk et al.~\cite{zaadnoordijk_can_2018}  and to Kelso \& Fuchs~\cite{kelso_coordination_2016}, we wanted to include the idea that the infant is attempting to predict the outcome of its actions in our model. Similar to Butko \& Movellan~\cite{butko_detecting_2010}, we assume this is an elementary aspect of infant interaction with the world that underlies the search for causation and agency. Within the context of current models of intrinsically motivated learning \cite{baldassarre2013intrinsically}, the approach we used could be said to be \textit{knowledge-based} rather than \textit{competence-based} and to involve predictive novelty motivation (cf.\cite{oudeyer2007intrinsic}). The central element of our model is a neural network and a table of the possible limb activities and their degree of ``\textit{interest}''. From that table, at each iteration of the simulation, the neural net chooses and attempts to execute the \textit{currently most interesting} limb activity and, at the same time, makes a prediction for the effect that this limb activity will produce on the mobile. When the prediction is wrong, this generates \textit{surprise}, and the degree of \textit{interest} of the issued command is increased in the interest table. If the prediction is right, the level of \textit{interest} for that command decreases in the table. Note that this scheme is distinct from reinforcement learning \cite{sutton2018reinforcement} where an agent seeks to maximize a sum of future and past rewards. In our model, there are no external rewards. The importance of prediction of action outcomes in capturing the causal nature is emphasized in the recent proposal of ``empowerment'' \cite{goddu2024development}
which aims to maximize the mutual information between the actions and the sensory outcomes.

The second guiding principle in our model was to include a motor system that was biologically more realistic than that used by Zaadnoordijk et al. \cite{zaadnoordijk_can_2018}. We wanted to simulate that in real infants, narrowing down actions to the particular limb that is connected to the mobile presumably represents a complicated problem. For that reason, our model assumed a realistic motor control mechanism whose effects on sensory feedback the infant has to learn.

To train the neural net, our model optimizes its functioning by minimizing three differences (referred to as \textit{losses}): the difference between its current activity and a baseline level of activity, the difference between the executed and the most interesting limb activity, and the difference between the predicted sensory feedback and the actual sensory feedback see equation \ref{lossesZ} in subsection \ref{NeuralNetworks}). The labels for these losses are generated at each timestep based on the actions of the model, so the network is trained online rather than with batches. Due to this setup, we do not have any sort of predefined datasets.

We shall see below that our model broadly reproduces infants' behavior. It quickly learns to differentiate and selectively increase the activity in the connected limb as compared to the unconnected limbs. The model also reproduces the control experiments showing that mere activity of the mobile does not suffice to produce increased limb activity: a proper contingency must exist between the infant's movements and the activation of the mobile. Our simulations also predict better contingency detection in the binary versus the conjugate condition, and provide insights into the controversial extinction burst. 

In addition to these main results, we perform several ``ablation'' studies, where we observe how the model performs when its main components are selectively removed (we will call the unablated model \textit{(complete model)}. These studies demonstrate that most of the components of the model are necessary ingredients: numerous muscles, motor noise, and the exploration mechanism attempting to minimize prediction error.

\section{Methods}

In this section, we describe how, in our model, we simulate the infant body, the environment, and the learning mechanisms. For this purpose, we start by introducing background information about the infant body and motor control. Then we describe the simulated mobile paradigm. Finally, we detail how we modeled infant learning and how our simulations match the number of participants and durations in infant studies \cite{jacquey_popescu_2020}, \cite{popescu_6-month-old_2021}.
\subsection{Body}\label{body}
Human motor control is highly complex and involves the interplay of a hierarchy of neural connections---cortical, subcortical and spinal---and a complex structure involving approximately 600 skeletal muscles \cite{edwards_motor_2010}, composed of muscle fibers, with a complex geometry (some acting over multiple joints) and nonlinear properties. The structures and their control show some redundancy---different motor commands can bring about the same movement; same effector movement can be instantiated through different joint configurations. We believe that the complexity and redundancy of this control mechanism is an important feature that shapes sensorimotor learning. In a real infant, through neural, mechanical or inertial connections, activating one muscle can affect multiple limbs. Direct effects arise because nerve fibers may innervate single or multiple muscles that affect the limb and even other limbs through multi-limb muscle-chain synergies. Indirect effects occur because limbs have masses and the body is in contact with the ground, and resultant inertial interactions mean that moving one limb may affect other limbs or the whole body.

Copying the entirety of the human motor control would have increased the complexity of our model beyond our current scope. Instead, we simplify the elements used in human motor control and abstract the entire chain by using a population code of 600 neurons constituting the muscle commands. In the model these muscle commands are transmitted via 600 output neurons of the neural network. The muscle commands are mapped by a function M to the four limbs of our simulated agent. This mapping works in an overlapping fashion across the limbs---a single command can have an effect on several limbs and limbs are controlled by multiple muscle commands. 

Also in order to simplify, we used a single scalar value between 0 and 1 as a proxy for each limb’s activity. This general definition of 'limb activity' allows us to make the link with data in the experimental literature where various measures of limb activity were used, for example, the frequency of kicks in Rovee \& Rovee \cite{rovee_conjugate_1969} or the change in limb acceleration over time in Popescu et al. \cite{popescu_6-month-old_2021}.

\subsection{Environment}

In a typical mobile paradigm experiment with infants, a limb movement may produce a sound or a visual change from the mobile. To model this, we represent the sensory feedback received by the model as a single value: either 0 or 1 in the binary condition, and a number between 0 and 1 in the non-binary (conjugate) condition (see next paragraph). For simplicity, in our simulation, we excluded all sources of sensory input to the agent except the mobile.

As in the mobile paradigm, we assumed that only one of the four limbs of the agent was ``connected'' to the mobile. For comparison with the empirical work of Popescu et al.\cite{popescu_6-month-old_2021} we explored two variants of the relation between the connected limb activity and the sensory feedback the mobile produces: a binary relation and a conjugate (or non-binary) relation. 
 
For the binary relation, when the activity of the connected limb exceeded a threshold, the mobile was activated for two iterations of the simulation. For compatibility with Popescu et al. \cite{popescu_6-month-old_2021}, where the threshold for triggering the mobile activation was increased in a few steps before reaching its maximum value, we did the same in our simulation (raising it from 0.5 to 0.6 in increments of 0.02). 

For the conjugate (or non-binary) relation, the intensity of the sensory feedback emitted by the mobile was directly proportional to the value of the connected limb activity. 

After the contingent phase of the experiment where the mobile response was present, we simulated the extinction phase included in some empirical work (e.g. \cite{zaadnoordijk_movement_2020, heathcockPerformanceInfantsBorn2004}; for a review, see \cite{bednarski_infants_2022}) by removing sensory feedback from the mobile for the last steps of each run.
\begin{figure*}
    \centering
    \includegraphics[width=\textwidth]{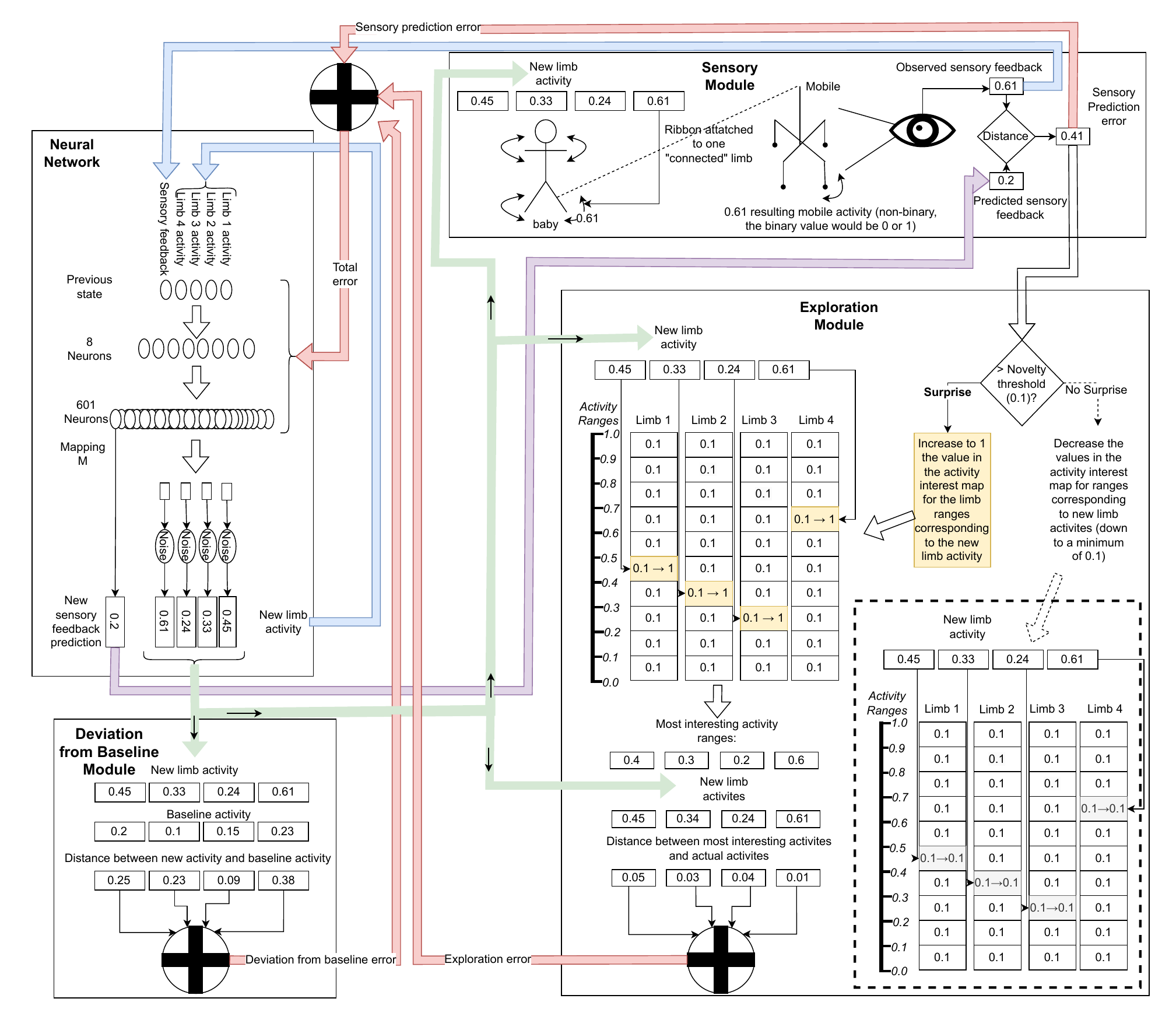}
    \caption{\textbf{Architecture of our model.} The architecture shows the four components of our model. The red arrows show the aspects of the model that have a direct influence on the network’s weights (the way the errors are computed and combined is defined in the Loss equation \ref{lossesZ}). The green arrows depict the flows related to the new activity of the limbs. The purple arrow shows the flow of the new sensory feedback prediction and the blue arrows show the flows of the input data of the neural network. We separate the model into four components, on the top left is the Neural Network (see section \ref{NeuralNetworks}), on the bottom left is the Deviation from Baseline Module (see section \ref{baslineDev}), on the bottom right is the Exploration Module (see section \ref{exploration}) and on the the top right is the Sensory Module (see section \ref{sensoryFeedback}). In the Exploration Module (bottom right), we show the way the limb activity is processed; it is further explained and exemplified in Fig. \ref{fig:AIM} and detailed in subsection \ref{exploration}. In the Neural Network part (top left), the mapping M is exemplified in Fig. \ref{fig:BetaDistributions}.}
    \label{fig:Flowchart}
\end{figure*}

\subsection{Learning Mechanism}
 As shown in Fig. \ref{fig:Flowchart}, our learning mechanism consists of four modules: the neural network, a sensory prediction module, an exploration module, and a deviation from baseline module (the inner workings of each module are detailed below in separate subsections). Its purpose is to control the system's behavior over successive time steps in order to take actions and make predictions about the sensory effect that it expects from those actions. On the one hand, the mechanism tries to continuously improve the predictions of the sensory feedback, therefore reducing the amount of surprise it experiences. On the other hand,  the mechanism also chooses actions that are likely to result in novel experiences, in this way making surprises more likely. The result of these two opposing forces is to make the model undertake a continuous non-random exploratory activity, seeking out the most surprising actions it can perform. 
\subsubsection{Neural network} \label{NeuralNetworks}

At each step of a simulation, the neural network takes the previous state of the system as its input and generates both new limb activities and a prediction for the sensory input that the network expects to receive. The previous state is represented by four neurons with activations between 0 and 1 corresponding to the activity of each of the four limbs and a fifth neuron with activation between 0 and 1 corresponding to the sensory feedback received from the environment. The five input neurons are fully connected via learnable weights to 8 hidden neurons, which are in turn fully connected via learnable weights to 601 output neurons (or fewer in the ablation studies). 600 of these correspond to the muscle commands (see Section \ref{body}), and one corresponds to the predicted sensory effect.

\begin{figure}
    \centering
    \includegraphics[width=\linewidth]{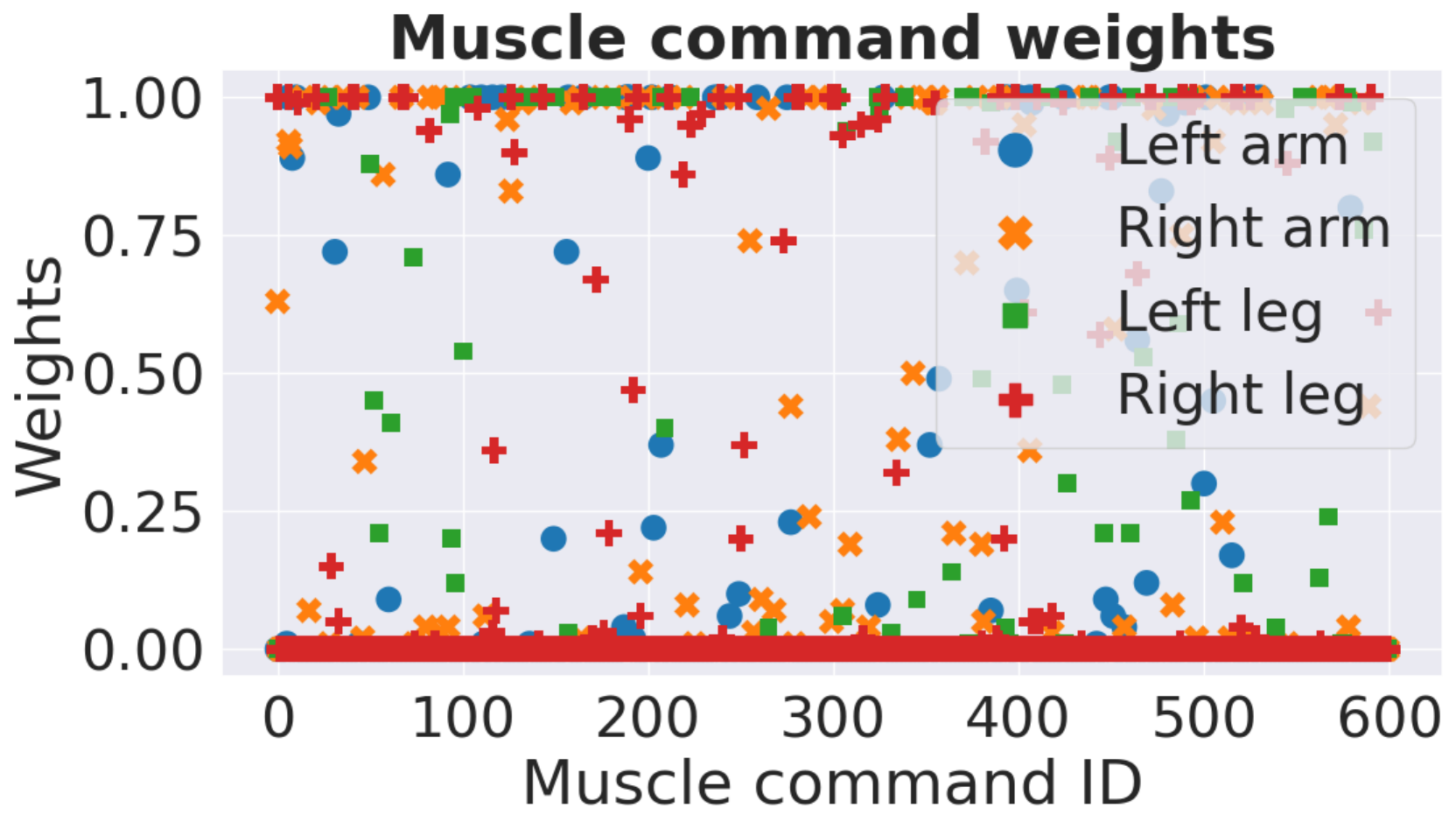}
    \includegraphics[width=\linewidth]{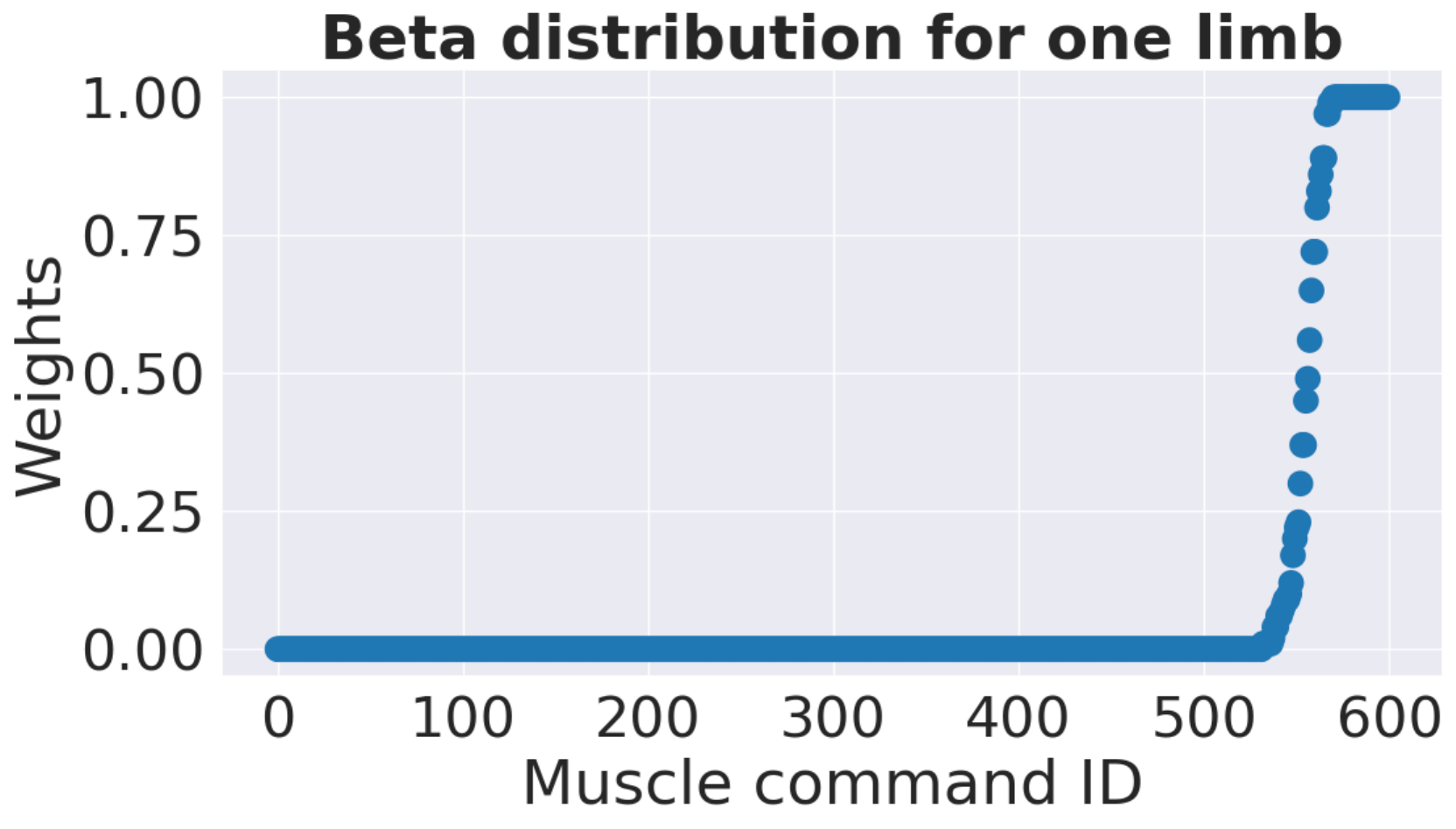}
    \caption{\textbf{Plots of example distributions of the used beta function.} 
    We used the beta function with parameters $z_1 = 0.01$, and $z_2 = 0.1$ to instantiate the particular connectivity between muscle command neurons and muscles that we desired in the model. For a particular limb, this distributed mode of command attributes a weight between 0 and 1 to each muscle command neuron. The weight defines the influence of the neuron for that limb. From our beta function, there exist ~25 neurons with a weight of 1 for a given limb, ~25 neurons with a weight between 0 and 1, and ~550 neurons with a weight close to 0. Any muscle command neuron has four weights in total, one for each limb, allowing for a single neuron to influence multiple limbs. This is shown in the upper panel of the figure, which shows all muscle command neurons unordered by weight, color-coded separately for each limb (see legend). The lower panel shows the weights reordered and sorted by weight for a single limb so that the distribution of weights for a single limb becomes clearer (here for the left arm). 
Muscle command neurons unordered by weight, color-coded separately for each limb
(see legend). (Lower) An example of muscle command neurons ordered by weight
for the left arm.}
    \label{fig:BetaDistributions}
\end{figure}

 We sought a mathematical function that instantiated the kind of connectivity between motor neurons and muscles that was inspired by the distributed, population codes used in humans. We assumed that there were 600 muscles in the whole body and these were controlled using a population code: Specifically, to approximate human anatomy, we wanted to have a distribution where ~25 neurons have direct control  (a “weight” of 1) for a limb and ~25 have a lesser effect (a “weight” between 0 and 1). The idea behind this is that there can be direct controlling effects - contracting muscles directly in the arm - or indirect controlling effects - contracting muscles in the chest, which also have slight effects on the arm. Furthermore, it should be possible to have neurons with overlapping control for multiple limbs. Many mathematical functions satisfy these characteristics. A convenient choice is the beta function with parameters $z_1 = 0.01$, and $z_2 = 0.1$ fulfills these requirements. We attribute four weights to each of the 600 neurons in the muscle command, each weight determining the degree to which that neuron influences each of the four limbs (in the model, this is expressed as four arrays with 600 values each). We sample the four weights from a distribution between zero and one, such that on average about 550 out of 600 samples would be near 0, 25 out of 600 samples would be near one and 25 would be between zero and one. The beta function (shown in Fig. \ref{fig:BetaDistributions}):

\begin{equation}
    B(z_1,z_2) = \int_{0}^{1}t^{z_{1}-1} (1-t)^{z_{2}-1} dt
\end{equation}
with $z_1 = 0.01$ and $z_2 = 0.1$ has this property.

Fig.~\ref{fig:BetaDistributions} (top) shows an example of attribution of weights of each of the neurons in the muscle commands to the four limbs following this scheme. Fig.~\ref{fig:BetaDistributions} (bottom) reorders the neurons of the muscle commands in order of increasing activation for the right arm only, to show that we have obtained the desired connectivity. Muscle command neurons 1 to 550 have a weight of zero, meaning they do not influence the right arm. Muscle command neurons 550 to 575 have intermediate weights and muscle command neurons 575 to 600 have weights of 1, meaning they strongly activate the right arm. Similar findings would apply to the other limbs but for different muscle commands. Note that during each run of the model corresponding to an experiment with one infant, the weights linking muscle commands to limbs stayed the same. To simulate different infants from infant experiments \cite{jacquey_popescu_2020}, \cite{popescu_6-month-old_2021}, the weights were resampled randomly for each run.

After the 600 muscle command neurons have been mapped to their respective limbs, the resultant total limb activity is calculated for each limb and motor noise is added. We chose a wide range of motor noise, specifically a random value between -0.3 and 0.3, in order to increase the difficulty of the task and enforce some amount of exploration through variability. Finally, after the generation of muscle command neurons' activations, their mapping to the limbs, and the addition of the motor noise, the network has produced the new limb activity.

All layers in the network are fully connected linear layers, meaning that each neuron in one layer is connected via a particular weight to each neuron in the previous and succeeding layers. Linear layers apply a linear transformation to incoming data: $y = xA^T + b$, where y is a vector of output values for the next layer, x the vector of input values for the layer, $A^T$ is a matrix of weights and b is a vector of the biases of each neuron in the layer. The weights and biases in each layer are learnable parameters of the model. The weights and biases are constantly adjusted in order to optimize a defined loss. This adjustment is done through the so-called backpropagation, which means that the loss is propagated through each connection and neuron of the model and adjusts weights and biases in a way that will reduce future losses. This means that if the model receives the same input as before, it will compute an output that is slightly different from the previous one and that would be closer to the output expected by the loss function.

The loss function of our model consists of three terms, each specifying a distance using the mean square error (MSE). The first of these terms evaluates the accuracy of the predicted sensory feedback. The second term encourages exploration within the model. The third term ensures that the activities of the limbs remain close to a preferred baseline level. The three terms are combined as shown in the following equation:
\begin{equation*} 
    \textit{Loss} =\  \textit{MSE} (\textit{predicted\_sensory\_feedback}, \textit{sensory\_feedback})
\end{equation*}
\begin{equation*}
     + \textit{ MSE} (\textit{activity}, \textit{desired\_activity})
\end{equation*}
\begin{equation}
\label{lossesZ}
     + \textit{ MSE} (\textit{activity}, \textit{baseline\_activity})
\end{equation}

To simulate that there were different infants in the behavioral experiments, the weights and biases are initialized randomly for each simulation. Each simulation begins with an initialization period of 1000 steps without any contingency in order to force the model’s baseline activity to approximate real infants’ starting activity. Based on infant data of previous studies \cite{popescu_6-month-old_2021}, \cite{jacquey_popescu_2020}, we chose an activity with a mean of 0.15 and a standard deviation of 0.15 for the baseline. Time “0” of each run then starts after the initialization period.

\begin{figure*}[ht]
    \centering
    \includegraphics[width=1\linewidth]{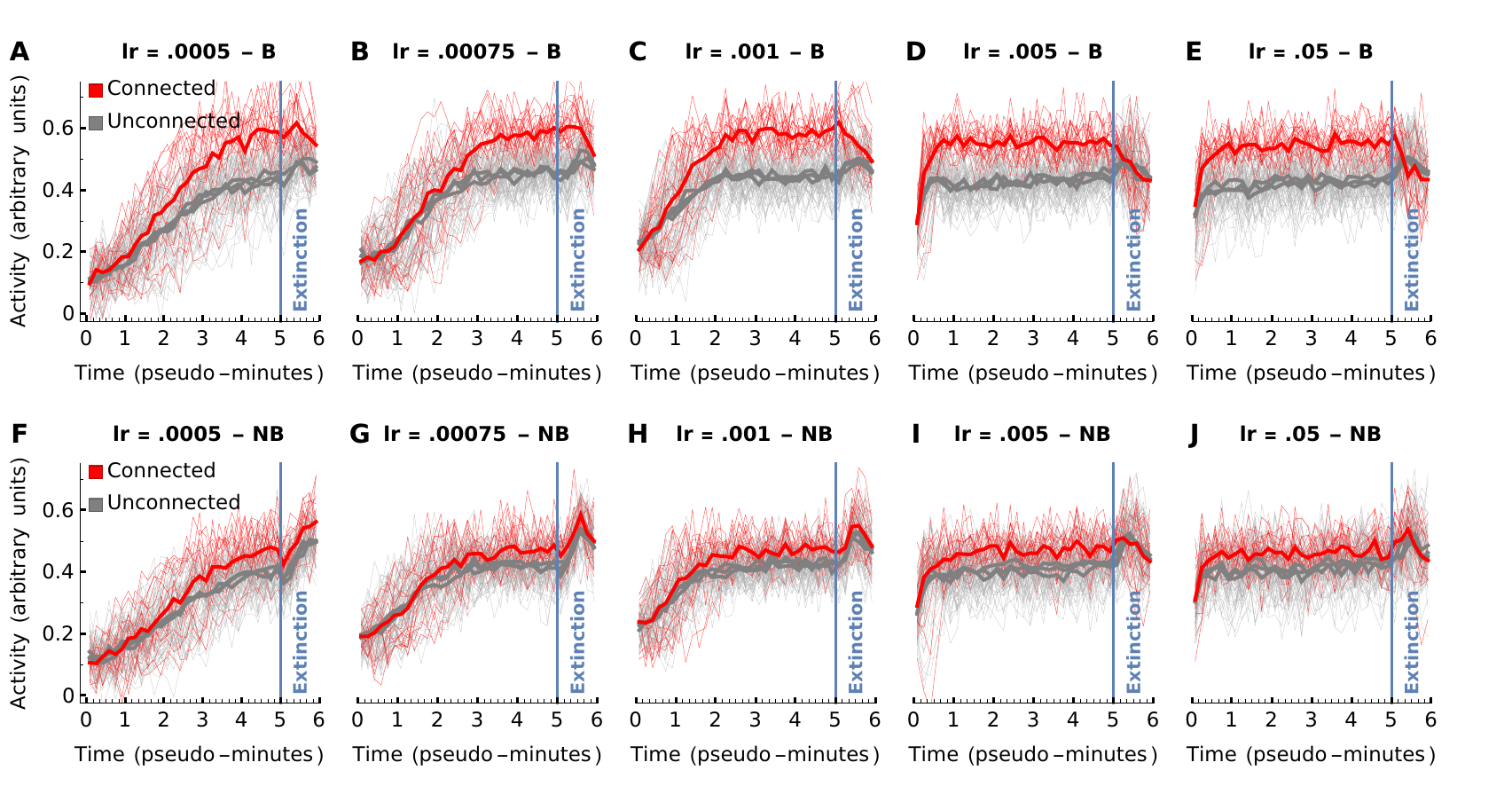}
    \caption{\textbf{Models with different learning rates.} The thick curves show the mean activities, across 20 individual simulations runs, of the connected limb (red) and the unconnected limbs (gray). Thinner pale curves show 20 individual simulation runs. Note that our complete model has a learning rate of 0.00075. (A, B, C, D, E) Binary simulations with learning rates of respectively 0.0005, .000075, .001, .005, and .05.  (F, G, H, I, J) Non-Binary simulations with learning rates of respectively .0005, .00075, .001, .005, and .05.}
    \label{fig:LearningRate}
\end{figure*}

The learning rate of our model is another important parameter determining the results and perhaps the parameter hardest to translate to the experiments in the real world. For infants, a higher learning rate could mean that they are able to learn more quickly. In neural networks, the learning rate is generally defined as the amount by which the network’s weights are changed at each learning step. A higher learning rate means that each learning step has a higher impact. However, a compromise in learning rate must be reached because a learning rate which is too high often gives rise to oscillatory behavior instead of converging to a dynamic equilibrium, while a very low learning rate would mean that the model would take a long time to find the equilibrium. We used a constant learning rate of 0.00075 for all simulations. We chose this value after empirical testing and found that it gave results that best matched those of infant studies \cite{popescu_6-month-old_2021}, \cite{jacquey_popescu_2020}. Fig.~\ref{fig:LearningRate} shows examples of results with other learning rates.

\subsubsection{Sensory feedback}\label{sensoryFeedback}
The sensory feedback module has two purposes: (1) the simulation of the mobile and (2) the calculation of the prediction error. The simulated sensory feedback produced by the mobile is based on the activity of the connected limb. The sensory feedback is a single scalar value that might represent any sort of sensory feedback used in experimental work, be it auditory or visual.

In our model, out of the four limbs, only one limb is connected to the mobile. In the binary condition, when the activity of the connected limb exceeds a threshold, the mobile is activated for two steps. In the non-binary (conjugate) condition, the intensity of the sensory feedback emitted by the mobile is directly proportional to the value of the connected limb activity. After the contingent phase of the experiment comes the extinction phase. In this phase, the connected limb was disconnected from the mobile (in practice, we removed the sensory feedback from the mobile). The extinction phase lasts for the last 240 steps of each run.

In addition to activating the mobile as a function of limb movements, the sensory feedback module must calculate the error that the neural network has made in predicting the sensory feedback produced by the mobile. This error is defined as the mean squared error between the predicted sensory feedback and the actual sensory feedback. This value is then fed back into the neural network, where it acts as another loss used to adjust the network's weights to improve its predictions.

\subsubsection{Exploration}\label{exploration}
The exploration module uses the sensory prediction error calculated by the sensory feedback module and the new limb activities from the neural network to update the \textit{activity interest map}. It also calculates an ``Exploration Error'' used as one of the losses to modify the weights of the neural network, guiding it to a more efficient exploration.
\begin{figure}
    \centering
    \includegraphics[width=\linewidth]{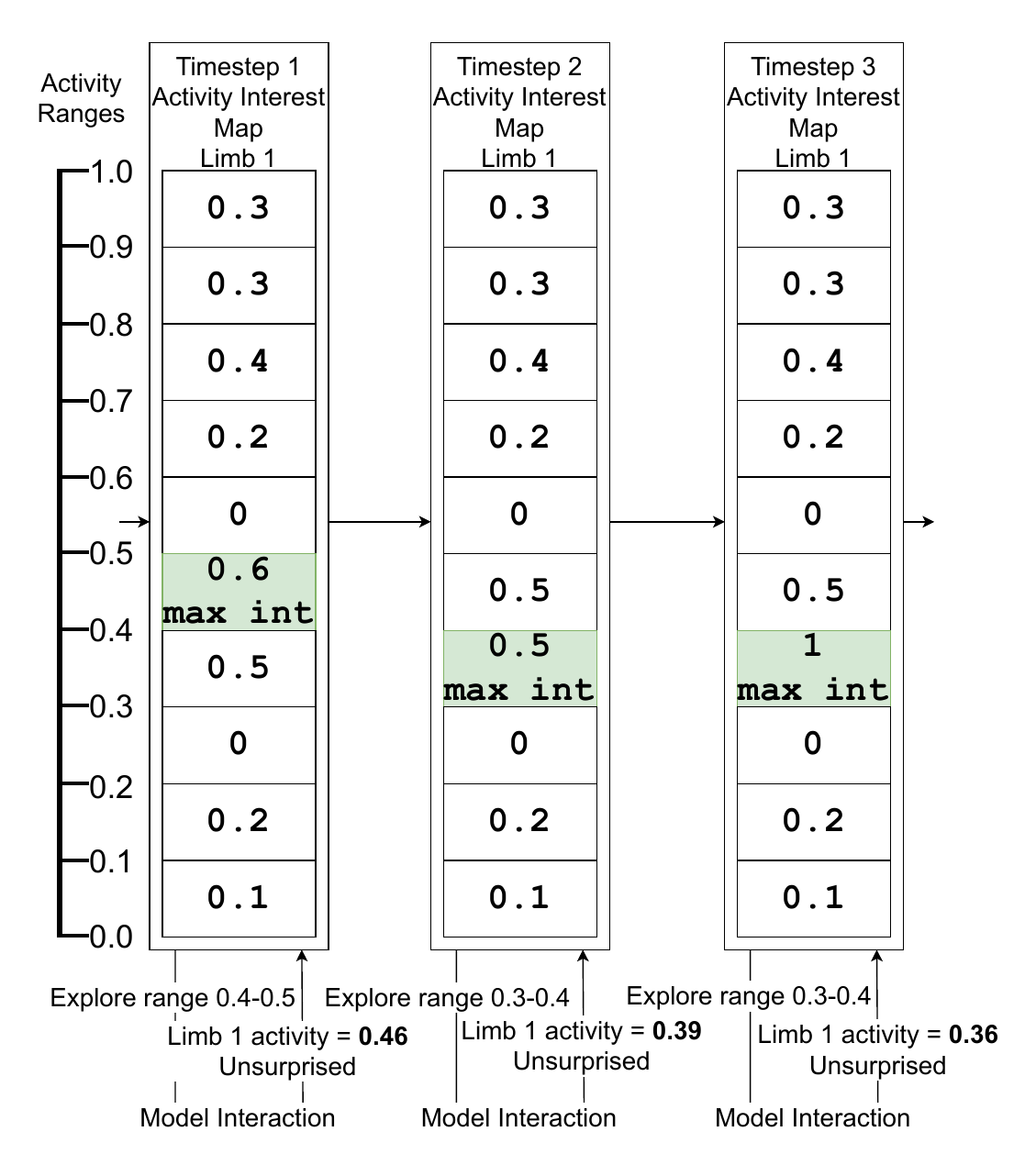}
    \caption{\textbf{An example process of three timesteps and corresponding changes in the activity interest map of limb 1.} 
    An example of three timesteps and corresponding changes in the activity interest map of limb 1. The activity is divided into 10 ranges: 0 to 0.1, 0.1 to 0.2, up to 0.9 to 1.0 Each range can have a different amount of interest from \texttt{\textbf{0}} to \texttt{\textbf{1}} (shown in \texttt{\textbf{typewriter bold characters}} to differentiate them from values related to activity). Assume that at Timestep 1, the activity range 0.4-0.5 happens to have an interest value of 0.6 that is the highest value currently in the interest table (shown on green background and indicated with \texttt{\textbf{max int}}). Assume that given the previous limb activity, and based on its current weights and biases the model at Timestep 1 produces an output limb activity of 0.46. This activity results in some sensory feedback. Let’s assume that the model correctly predicts this feedback and hence that no surprise is generated. Therefore, in Timestep 2 the interest value of the activity range 0.4-0.5 is reduced from \texttt{\textbf{0.6}} to \texttt{\textbf{0.5}}. As a consequence in Timestep 2 the highest interest value among all activity ranges is \texttt{\textbf{0.5}} and there are two activity ranges with that interest value, namely ranges 0.4-0.5 and 0.3-0.4. When there is more than one activity range with the highest interest value, one of them is randomly chosen. In this example this is the range 0.3-0.4 which is highlighted in green. Assume now that in Timestep 3 the model produces an activity in this range, 0.39, and is surprised by the sensory feedback. Due to being surprised, in Timestep 3 the interest in this range is raised to \texttt{\textbf{1}}.} 
    \label{fig:AIM}
    \vspace{-0.4cm}
\end{figure}
The activity interest map keeps track of two elements: 1) which limb actions ranges the model has explored before and 2) the extent to which it can predict the resulting sensory feedback for those ranges. When the model successfully predicts the sensory feedback resulting from an action, it should become less interested in the corresponding action range. However, when the sensory feedback is unexpected, we want the model to become highly interested in that action range. To accomplish this, we use an Activity Interest Map array to record whether the agent is ``surprised'' (see following paragraph) by the sensory result of its previous limb action. We discretize the continuous values for each limb activity into 10 action ranges of equal size, the first one for action values between 0 and 0.1 and so on, resulting in an overall array of dimensions (4,10) (four limbs and 10 ranges for each). At each simulation step, if a limb action generates surprise, the value in the table for the corresponding action range is directly set to 1, indicating high interest. However, if the limb action generates no surprise, the value in the array for the corresponding action range is decreased by 0.1, implementing a gradually decreasing interest. An example of the progress in the Activity Interest Map is shown in Fig. \ref{fig:AIM}. The array is initialized with values of 0.1 for each range. This can be interpreted as having a baseline level of interest in all possible actions. 

As the model's prediction is not binary, we introduce a ``Novelty Threshold'' that is set to 0.1. The model becomes ``surprised'' by the result of its action when the sensory prediction error generated by its limb action exceeds this novelty threshold. The novelty threshold determines the agent’s sensitivity to discrepancies between actual and predicted sensory feedback. 

In the exploration module, once the activity interest map has been updated as a function of whether surprise has occurred, the model must modify the neural net so that in the future the agent will tend to make actions within ranges of high interest. For this, the model calculates the mean square error (MSE) as the distance between the current action and the range of activity with the highest interest (should there be more than one range of equal interest, one of them is randomly chosen). This distance is used as a loss to adjust the weights of the neural network so that future muscle commands lead to an activity closer to the area of highest interest.

\subsubsection{Deviation from baseline activity}\label{baslineDev}
We incorporated a ``deviation from baseline'' module to emulate the fact that, on the one hand, infants presumably act to keep their activity in a restricted range around a certain baseline, but, on the other hand, they also tend to get globally more agitated, with that baseline increasing as the experiment proceeds. Deviation from this baseline is used as another loss that modifies the functioning of the neural network. 

The baseline is defined as four random values between 0 and a maximum value (set at 0.3 at the start), one for each limb, chosen at each simulation step. These emulate how each of the infant's four limbs would tend to move naturally in the absence of any sensory stimulation. To emulate that infants tend to grow fussier over time, the maximum value is increased by 0.0001 at every simulation step up to 0.454 at the end. The deviation from baseline is measured using the mean squared error between the current four baseline values and the current four limb activity values. Note that we assume that infants are not only fatigued by moving above their baseline activity but also that they are fatigued by keeping the same posture for a long time resulting in prolonged contraction of the same muscles.

\subsection{Number of simulation runs and their duration}
\subsubsection{Correspondence between the number of participants in the experiments and the number of simulations}
To parallel the number of infants in the two empirical studies, 20 in the binary condition in \cite{popescu_6-month-old_2021} and 18 in the non-binary condition in \cite{jacquey_popescu_2020}, we performed 20 simulations for each of the two conditions.

\subsubsection{Correspondence between experimental and simulation time}
When comparing the model outputs with real infant experiments, we must decide how to link the time steps of the simulation (i.e. each pass through the loop of the simulation) with time in a real experiment. In our simulation of the binary condition, the mobile activation lasts for 2 simulation steps. In the original study \cite{popescu_6-month-old_2021}, the mobile activation lasted 0.5 s. We took this as a reference and therefore decided to consider that a simulation step is equivalent to 0.25 s. Therefore, to simulate the duration of the original study of 300 s, the simulation included 1200 steps, or 5 \textit{pseudo-minutes}. To allow the model to reach the baseline activity level, we added 1000 steps without the mobile at the beginning of the simulation (excluded from data analysis). To check if there is an extinction burst, we also added 240 steps without the mobile at the end of the simulation. Therefore, an entire “run” of the simulation involved 2440 steps or loops through the model. This ensures that the durations of infant experiments and of our simulations are comparable.

\section{Results}
We start this section by checking whether our model replicates the two main empirical results, \textbf{Specificity of connected limb} and \textbf{Comparison with a control group}. We then look at what can be deduced from our simulations of the \textbf{Extinction burst}. Regarding the replication of the difference between the binary and conjugate stimulation, we report it within each of the first three subsections (we compare model simulation data with empirical data on 6-month-old infants separately for the binary and non-binary\footnote{Note that we show graphs for both binary and non-binary conditions for every result, we will discuss the difference between these in subsection \ref{binaryVsNonBinary}.}, taken respectively from ~\cite{popescu_6-month-old_2021} and \cite{jacquey_popescu_2020}) and conclude on in the subsection \textbf{Greater effect for binary versus non-binary}. In the last subsection, we consider the results of ablation studies where we do simulations in which we remove different components of the model to test which of them are essential for the results.

\subsection{Specificity of connected limb}
\begin{figure}
    \centering
    \includegraphics[width=\linewidth]{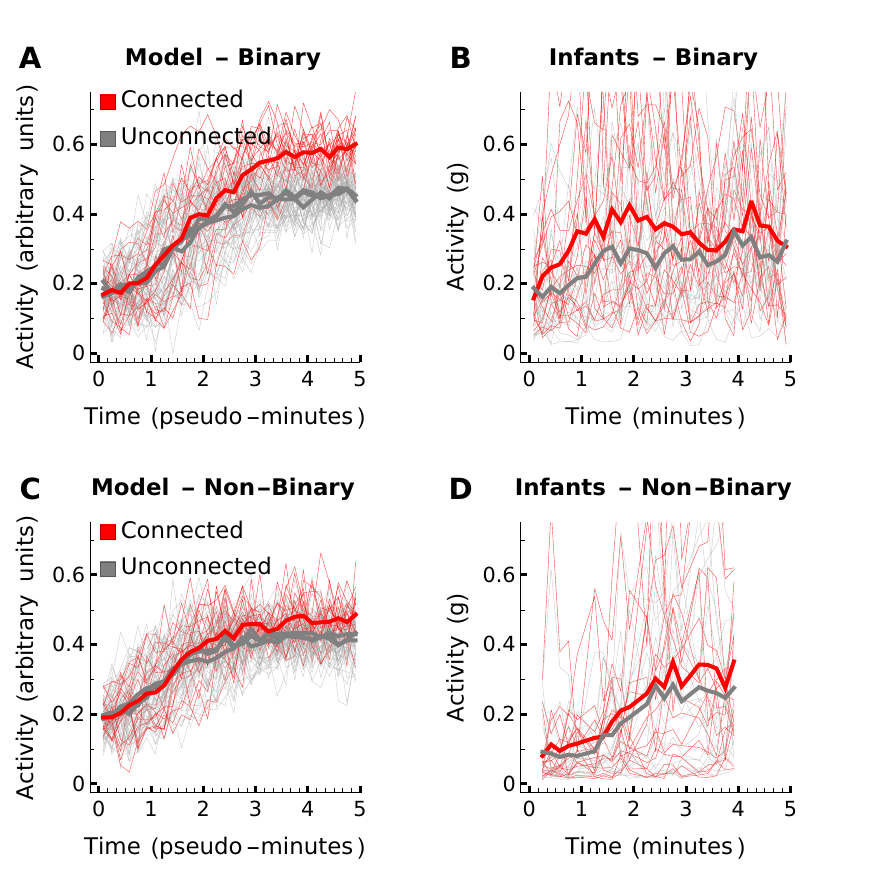}
    \caption{\textbf{Comparison of limb activity in simulations and infants (mean activity per limb and per 10-s bin).} Activity in simulations is expressed in arbitrary units and activity in infants is expressed in gravitation acceleration units. The thick curves show the mean activities across individual data, of the connected limb (red) and the unconnected limb(s) (gray). Thinner pale curves show individual data (individual simulation runs or infants). (A) Model in the Binary condition (20 simulation runs). (B) Infants in the Binary condition (data on 20 six-month-old infants in Popescu et al. \cite{popescu_6-month-old_2021}). (C) Model in the Non-Binary condition (20 simulation runs). (D) Infants in the Non-Binary condition (data on 18 six-month-old infants in Jacquey et al. \cite{jacquey_popescu_2020}; note that this experiment lasted only 4 minutes and that the data on the attention-getter corresponding to the first bin of every minute was removed).}
    \label{fig:SimVsInfant}
\end{figure}

Fig. \ref{fig:SimVsInfant}, shows the graphs of the moment-to-moment activity of the four limbs in the simulations and a comparison with the corresponding infant data of the two arms. The most important fact is that the model does what it was expected to do: like infants, it \textbf{is able to differentiate the connected and unconnected limb(s)}. However, the model is somewhat slower in doing this, requiring more than one minute, whereas infants seem to differentiate the limbs at the start of the first minute. 

An obvious difference between the model and the infants is that the overall variability of activity is much greater for the infants than for the model. Compared to the model, between-participant variability may be greater in infants because they vary more in intrinsic parameters like body morphology, degree of maturation, and learning rate. Within-participant variability in infants may also be greater because they suffer moment-to-moment variations in attention, changes in interest and distractions that we have not modeled in our simulation.

\begin{figure}[t]
    \centering
    \includegraphics[width=1\linewidth]{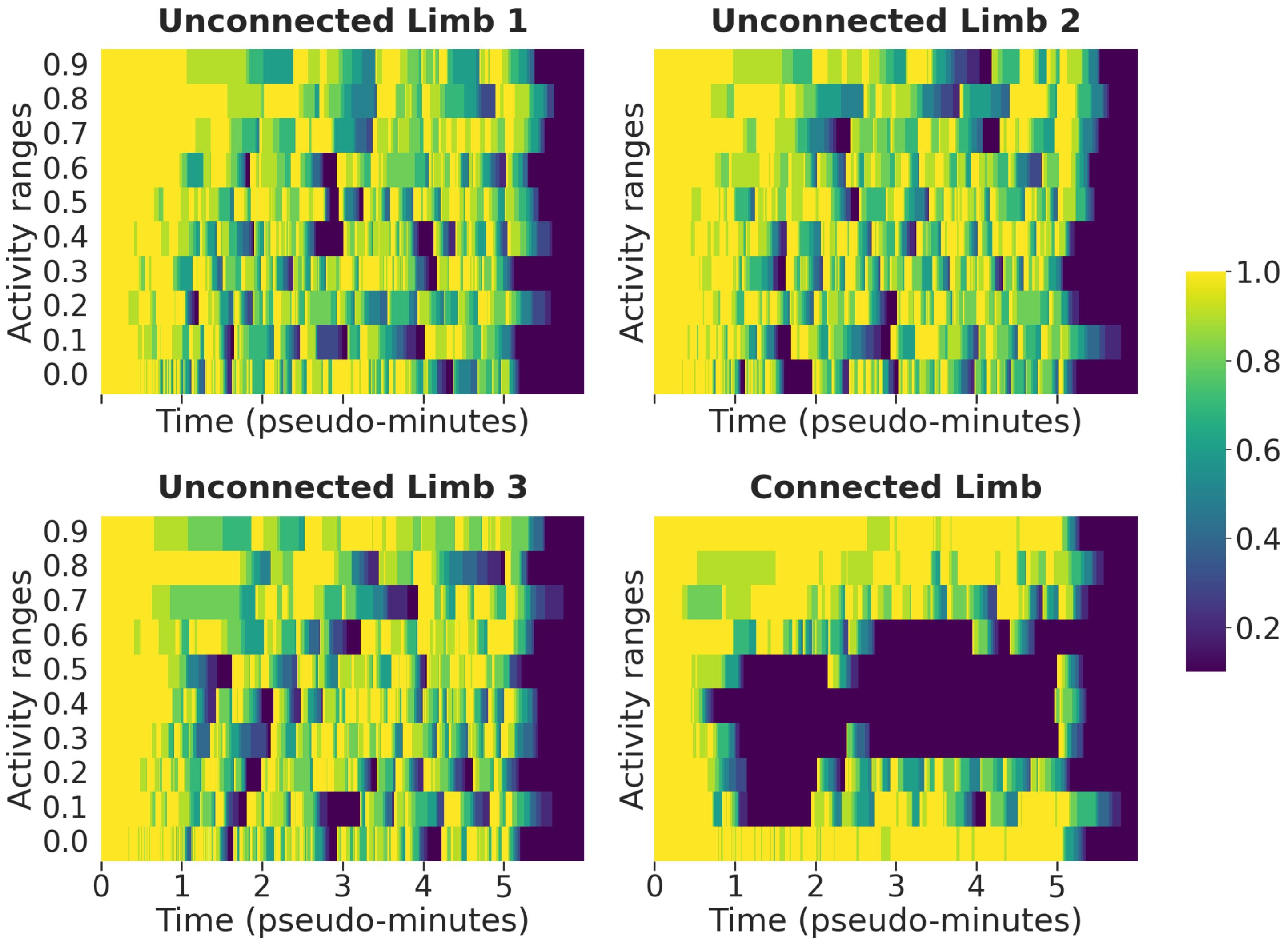}
    \caption{Evolution of the activity interest map for each limb, example of a run in the non-binary condition. For each limb, the figure shows how the values of interest for each activity range evolve over time. The X-axis shows the time and the Y-axis shows the activity ranges. The color of each (x,y) cell shows the interest of that activity range at that specific time. The blue-yellow spectrum represents the scalar value corresponding to the interest, blue color corresponds to lowest interest values, greenish color--to intermediary and yellow color--to highest interest values.}
    \label{fig:heatmapConUnCon}
    \vspace{-0.3cm}
\end{figure}

To provide a deeper look into the mechanics of the model and how it can identify the connected limb, let us zoom in on the Exploration module (see Fig. \ref{fig:Flowchart}), specifically on the Activity Interest Map. The main goal of this sub-module is to guide exploration in our model. It does so by calculating how interested the model is in each limb activity. Fig. \ref{fig:heatmapConUnCon} shows how the values of ``interest'' in the Activity Interest Map change over time for each limb. The difference between the connected (lower right panel) and the unconnected limbs is evident. This shows how the model identifies which limb is connected and explores its activity distinctively.

In Fig. \ref{fig:heatmapConUnCon}, the colors from blue to yellow represent rising values of ``interest", with blue corresponding to lowest interest, green shades to intermediate, and yellow to highest interest. For the connected limb, there is a clear separation between blue and yellow areas: blue areas (little interest) are found near where most of the limb's activity is occurring, namely near values of about 0.5 (see Fig. \ref{fig:SimVsInfant}). For activity ranges for which the model has learned to predict the sensory feedback, the model is no longer ``surprised'' and will explore them less than the very high and very low activity ranges. For the unconnected limbs, there is no statistically reliable link between the limb's activity and the sensory feedback, and so no reliable predictions can be made for any activity range; we see fairly high interest (yellow shades) for all activity ranges. The region in the heatmaps after minute 5 corresponds to the time when the contingency has been removed. The model always receives the same sensory feedback set at 0, and so becomes able to predict the results of all of its actions. There is no interest in any activity region.

\subsection{Comparison with a control group}
\begin{figure}[t]
    \vspace{-0.4cm}
    \includegraphics[width=1\linewidth]{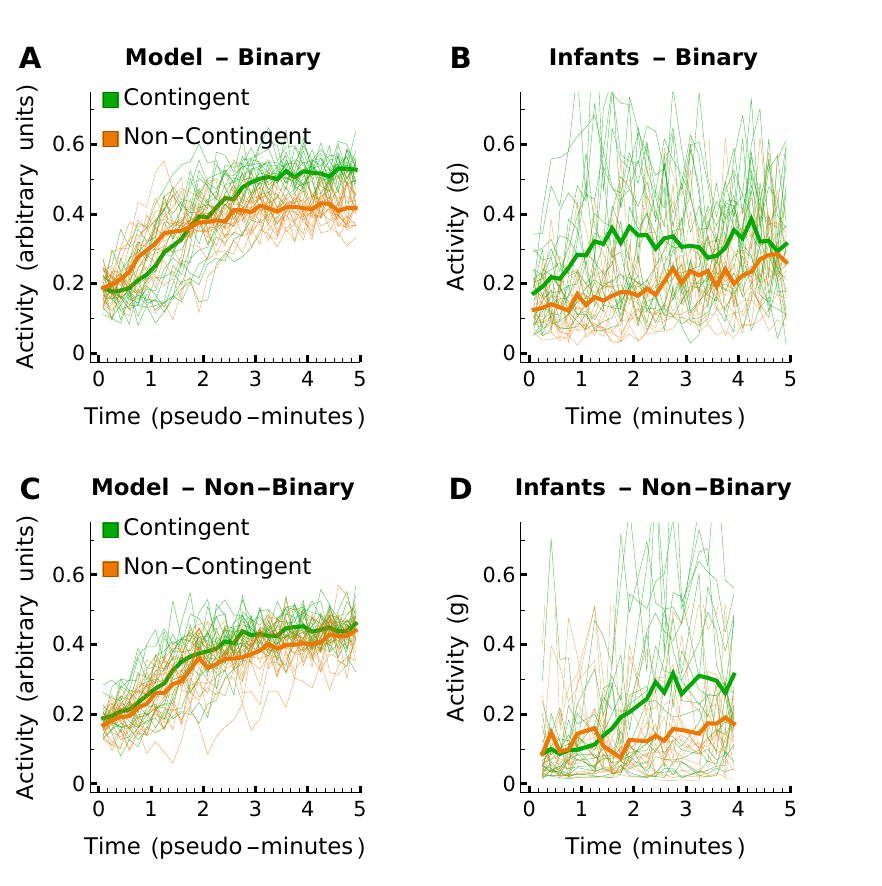}
    \caption{\textbf{Mean limb activity in Contingent and Non-Contingent conditions (mean over joint activity of the connected and unconnected limbs per 10-s bin).} Note that for consistency with infant data, in the complete model we selected one unconnected limb among the three available). The thick curves show the mean activities, across individual data, of the Contingent group (green) and the Non-Contingent group (orange). Thinner pale curves show individual data (individual simulation runs or infants). (A) Complete model in the Binary condition (20 simulation runs). (B) Infants in the Binary condition (data on 20 6-month-old infants in Popescu et al. \cite{popescu_6-month-old_2021}). (C) Complete model in the Non-Binary condition (20 simulation runs). (D) Infants in the Non-Binary condition (data on 18 6-month-old infants in Jacquey et al. \cite{jacquey_popescu_2020}); note that this experiment was shorter (4 minutes), and that, similar to the original study, the data on the attention-getter corresponding to the first bin of every minute was removed.}
    \label{fig:contingentvsNot}
\end{figure}

We not only want to verify that the model correctly simulates what infants do in the presence of contingent sensory feedback (Fig. \ref{fig:contingentvsNot} A and C), but also what they do when the sensory feedback is not contingent on their actions, that is, when the stimuli are triggered independently of limb movements. Experimental data obtained for infants in this case is presented in Fig.~\ref{fig:contingentvsNot} B and D. For the non-contingent condition, the stimuli from the contingent group were ``replayed". In this way, the frequency and intensity of stimuli in the contingent and non-contingent groups were equated. We did the same in our model. The most important finding is that our model correctly simulates the higher activity in the contingent group compared to the non-contingent (control) group, as shown in Fig. \ref{fig:contingentvsNot}. It indicates that, \textbf{like infants, the model is sensitive to the contingency itself, not just the sensory stimulation coming from the mobile.}

The difference between the contingent and non-contingent groups are somewhat smaller in the model than in infants. Also, for the model, the shape of the curves for contingent and non-contingent conditions is fairly similar, starting with a rising slope and then flattening off. On the other hand, for infants, the curves are overall flatter and may (in the case of the non-contingent condition) not have a rising part at all. This could suggest that in the non-contingent condition, infants realize from the beginning that they have no impact on the stimuli and so move less than in the contingent condition. The model, in contrast, ``believes'' for longer that it affects the stimuli and searches longer for a way to use its limbs to manipulate the sensory feedback. Another difference between the model and the infants is that the variability is much more significant in the infant data, similar to what we saw in the previous subsection.

Interestingly, there is a peculiar aspect of the curves for the binary condition, where at the beginning of the experiment the non-contingent group actually shows greater activity than the contingent condition, but at around 2 minutes into the experiment it falls below the curves for the contingent group. This behavior is unexpected, as there should be no reason for a higher activity in the non-contingent group at the start of the simulation.
\subsection{Extinction burst}

\begin{figure}[!t]
    \centering
    \vspace{-0.2cm}
    \includegraphics[width=\linewidth]{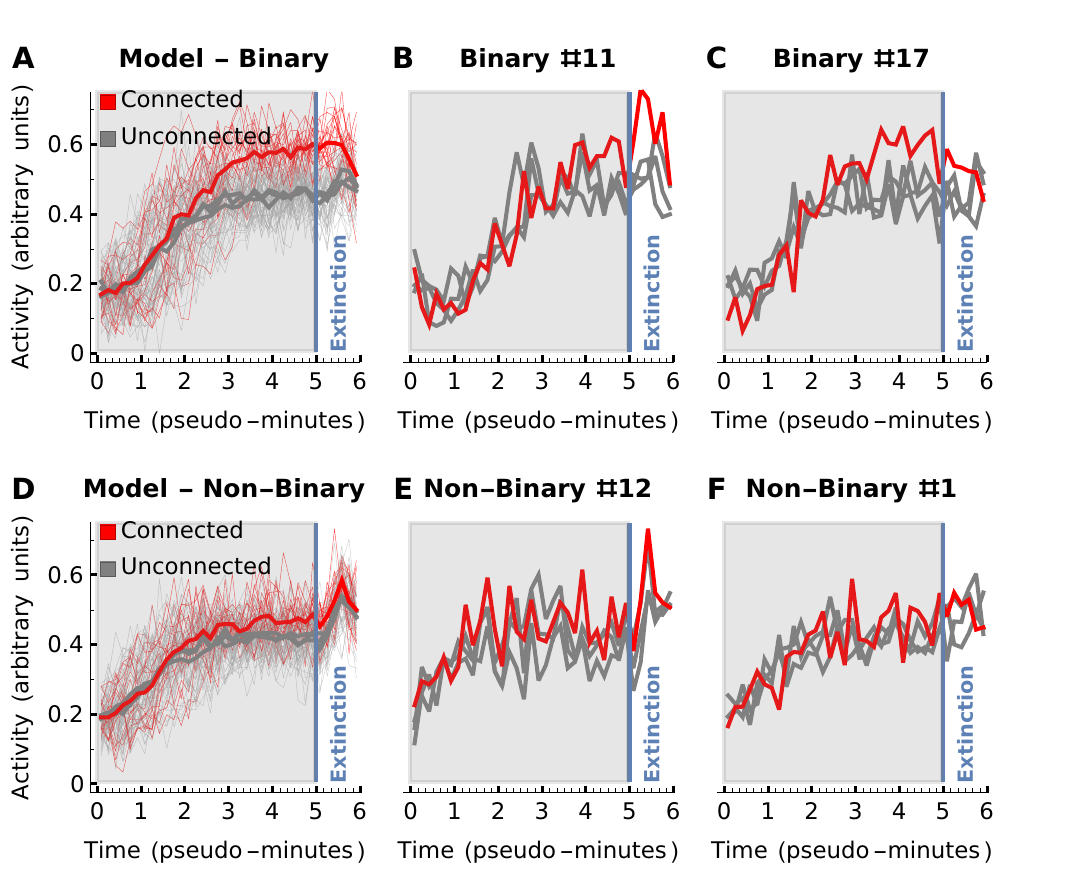}
    \caption{\textbf{Limb activity before and after removing the contingency (mean activity per limb and per 10-s bin).} On the Time axis, the moment of contingency removal is emphasized with a vertical line. The one-minute period after contingency removal is shown on contrasting background. (A \& D) The thick curves show the mean activities, across individual data, of the connected limb (red) and the unconnected limbs (gray). Thinner pale curves show data on individual simulation runs. (A) Model in the Binary condition (20 simulation runs). Examples of individual runs in the Binary condition (B) with an extinction burst and (C) without an extinction burst. (D) Model in the Non-Binary condition (20 simulation runs). Examples of individual runs in the Non-Binary condition (E) with an extinction burst and (F) without an extinction burst.}
    \label{fig:Extinction}
\end{figure}

In the literature on the mobile paradigm, an extinction burst is sometimes observed when, after the infant has learned to activate the mobile with the connected limb, this limb is disconnected from the mobile causing a sudden and transient burst of activity \cite{alessandri_violation_1990},\cite{heathcock_relative_2005},\cite{lewis_violation_1990},\cite{rovee-collier_topographical_1978}. Such extinction bursts can be interpreted as the reaction to a violation of expectations and loss of control \cite{lewis_violation_1990}. When in our model, we disconnect the connected limb, we also observe extinction bursts but they do not occur systematically, as shown in Fig. \ref{fig:Extinction}. This is similar to results on extinction bursts in infants (for a review, see \cite{bednarski_infants_2022}).

Fig.~\ref{fig:Extinction} shows the average results of our simulations along with two examples of individual runs. The vertical line in the graphs shows the moment of extinction of the contingency, marked as time 5 minutes. Whereas in the binary condition there seems to be no evidence for an extinction burst, there seems to be a clear one in the non-binary condition. The data for individual simulated infants are quite variable; two examples are shown in Fig.~\ref{fig:Extinction}~B-C and \ref{fig:Extinction}~E-F, one with a possible extinction burst, and one without an extinction burst. Additional plots are available in the Supplementary Material (Fig. S1, S2). In the individual data there does not seem to be clear evidence for a sudden increase in activity after the extinction of the contingency. Our findings are consistent with the fact that the existence of an extinction burst is debated in the literature, since different infants may have different learning rates, and depending on the sample, extinction bursts may be more or less evident.

\subsection{Greater effect for binary versus non-binary} \label{binaryVsNonBinary}
Our prediction for our model was that a binary contingency would be more easily learned and more effective in increasing the model's activity than a non-binary (conjugate) contingency. Indeed, this is what we find when comparing connected and unconnected limbs (Fig. \ref{fig:SimVsInfant} A and C) and contingent and non-contingent conditions (Fig. \ref{fig:contingentvsNot} A and C).
To explain these effects, we suggest that in the binary condition, a small change in activity can cause a large change in the sensory feedback, while in the non-binary condition, the changes in activity and sensory feedback are proportional. In the binary condition, a small limb activity that crosses the stimulus triggering threshold creates a strong change in the sensory feedback. Therefore, in the binary condition, it is easier to detect which is the connected limb, even when limb activity only varies slightly compared to other limbs. While in the non-binary condition, it is harder to detect the limb that is responsible for a small change in sensory feedback. Analogous differences between the two conditions are also observed in the ablation studies of the number of muscles commands and motor noise.

\subsection{Ablation studies}
To differentiate the unablated model from models with ablations we will further call the unablated model \textit{complete model}. Note that in all the ablation studies reported below the ablated models are trained in the exact same way as the complete model.
\subsubsection{No prediction error}
\begin{figure}[t]
    \centering
    \includegraphics[width=\linewidth]{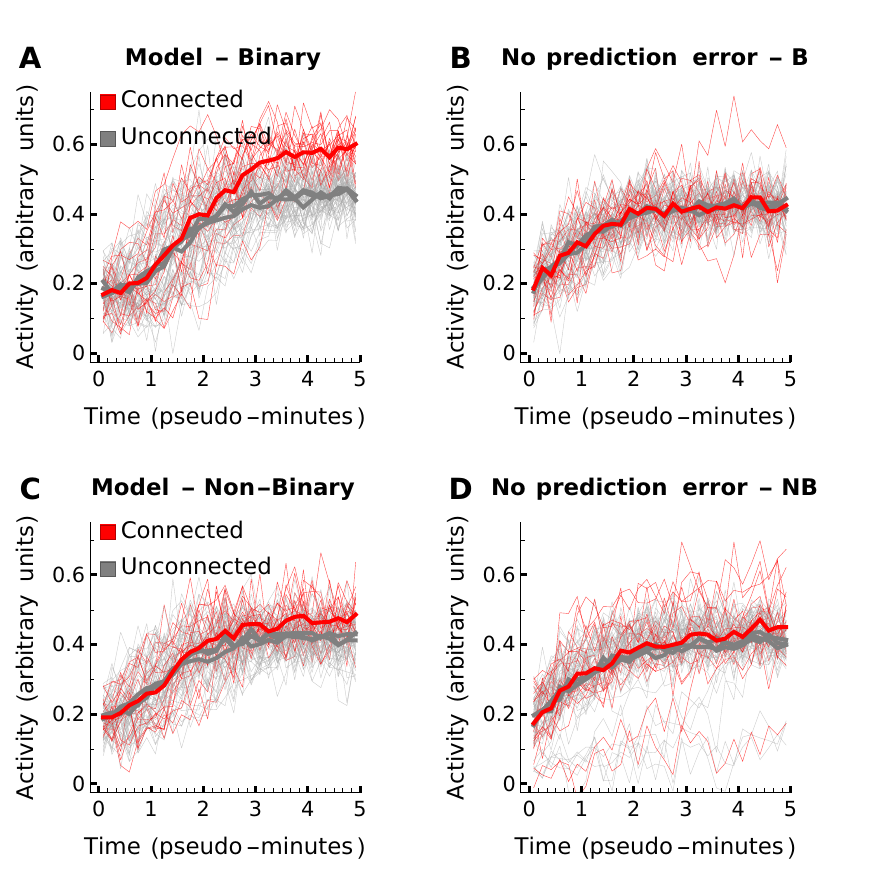}
    \caption{\textbf{Ablation of the prediction error.} The thick curves show the mean activities (mean activity per limb and per 10-s bin, data on 20 simulation runs), across individual data, of the connected limb (red) and the unconnected limbs (gray). Thinner pale curves show individual simulation runs. (A) Complete model in the Binary condition. (B) Ablated model in the Binary condition. (C) Complete model in the Non-Binary condition. (D) Ablated model the Non-Binary condition.}
    \label{fig:predictionAb}
\end{figure}

To see how the model works in the absence of the prediction error, we disconnect the prediction loss from the network, not allowing it to modify any of the weights. Therefore, this ablation shows how the model works when it cannot predict the effects of its actions. Fig. \ref{fig:predictionAb} shows the results of this ablation study. The effect of this ablation is to completely prevent the model from distinguishing connected and unconnected limbs, therefore, the model cannot learn to activate the mobile. Without the prediction loss, the model's behavior becomes mainly determined by which activity levels are labeled as causing surprise. Note that the model continues to compute ``surprise'' and modify the Activity Interest Map as a function of the prediction error. Therefore, in this case, once the model by chance finds a surprising activity, it continues to explore that activity range, with variations caused only by random motor noise. 
\subsubsection{No novelty-based reinforcement calculator}
\begin{figure}[t]
    \centering
    \includegraphics[width=\linewidth]{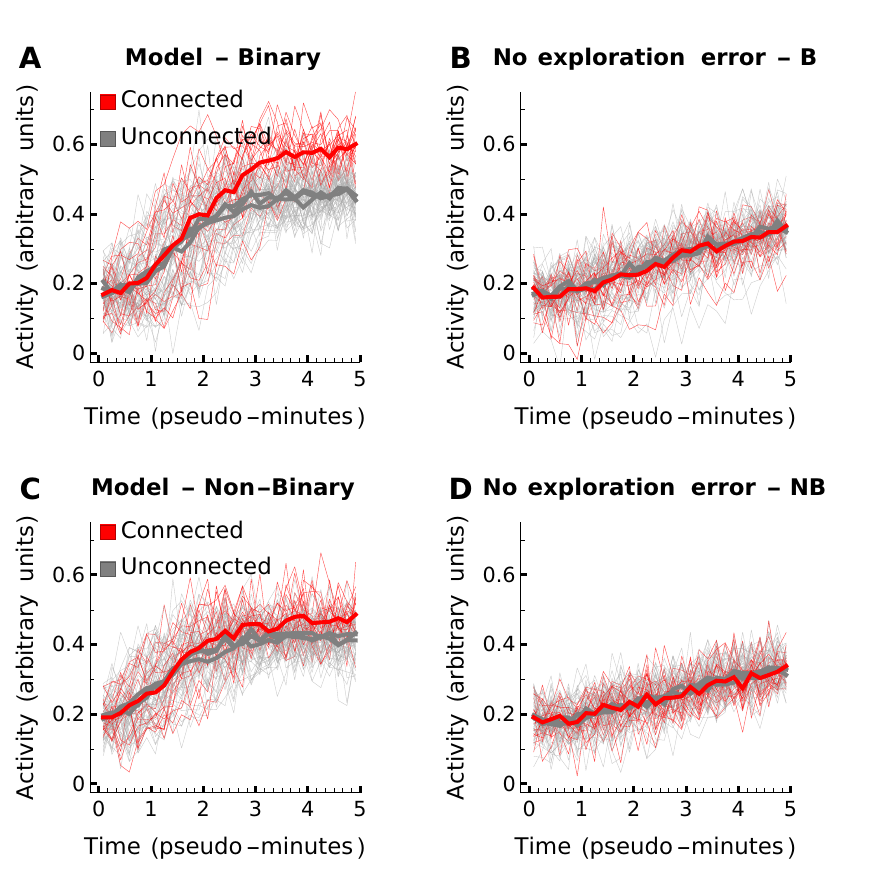}
    \caption{\textbf{Ablation of the exploration loss.} The thick curves show the mean activities (mean activity per limb and per 10-s bin, data on 20 simulation runs), across individual data, of the connected limb (red) and the unconnected limbs (gray). Thinner pale curves show individual simulation runs. (A) Complete model in the Binary condition. (B) Ablated model in the Binary condition. (C) Complete model in the Non-Binary condition. (D) Ablated model the Non-Binary condition.}
    \label{fig:expoAb}
\end{figure}

Fig.~\ref{fig:expoAb} shows the ablation run where the novelty-based reinforcement calculator is not used to modify the weights of the neural network. This ablation not only prevents the model from being able to distinguish between the connected and unconnected limbs but also lowers the overall limb activity. This effect is expected since the model has no incentive to explore, therefore it optimizes its prediction ability and reduces the deviation from baseline limb activity.
\subsubsection{Reducing the number of muscle command neurons}
\begin{figure}[t]
    \centering
    \includegraphics[width=\linewidth]{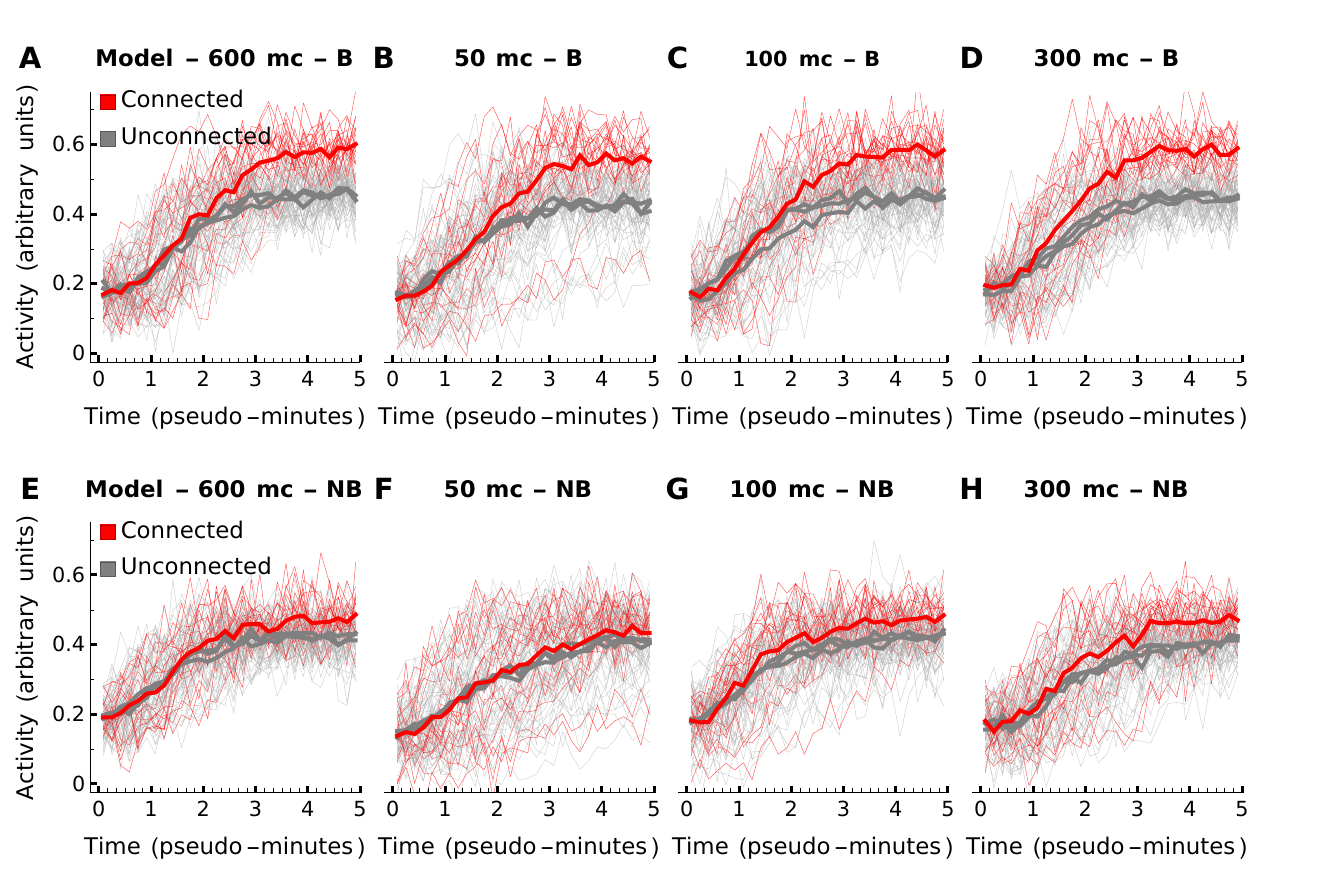}
    \caption{\textbf{Ablation of the output layer from 600 muscle command neurons to 50, 100 or 300.} The thick curves show the mean activities (mean activity per limb and per 10-s bin, data on 20 simulation runs), across individual data, of the connected limb (red) and the unconnected limbs (gray). Thinner pale curves show individual simulation runs. (A) Complete model in the Binary condition that includes 600 muscle command neurons (mc). (B, C, D) Ablation of the model in the Binary condition to 50, 100, and 300 muscle command neurons (mc), respectively. (E) Complete model in the Non-Binary condition that includes 600 muscle command neurons (mc). (F, G, H) Ablation of the model in the Non-Binary condition to 50, 100, and 300 muscle command neurons (mc), respectively.}
    \label{fig:muscleAb}
\end{figure}

In this ablation study, we changed the output layer from 600 muscle command neurons to 50, 100 or 300. The main finding is that, to consistently simulate infant behavior, the model needs a large number of muscle command neurons, as shown in Fig. \ref{fig:muscleAb}. The number of required muscle command neurons is larger in the non-binary condition (more than 300) compared to the binary condition ($\sim 100$).

The effect of this ablation is seen mainly in the variability across individual simulations. For small numbers of muscle command neurons, there are more individual runs in which the distinction between connected and unconnected limbs is not learned by the model (e.g. Fig. \ref{fig:muscleAb} B and F). In the binary condition, fewer muscle command neurons (100, Fig. \ref{fig:muscleAb}  C and G) are needed to reliably elicit a difference in activity between connected and unconnected limbs. In the non-binary condition, even with 300 muscle command neurons individual runs that do not learn to differentiate the limbs remain.

\subsubsection{Changing the amount of motor noise}
\begin{figure}[ht]
    \centering
    \includegraphics[width=\linewidth]{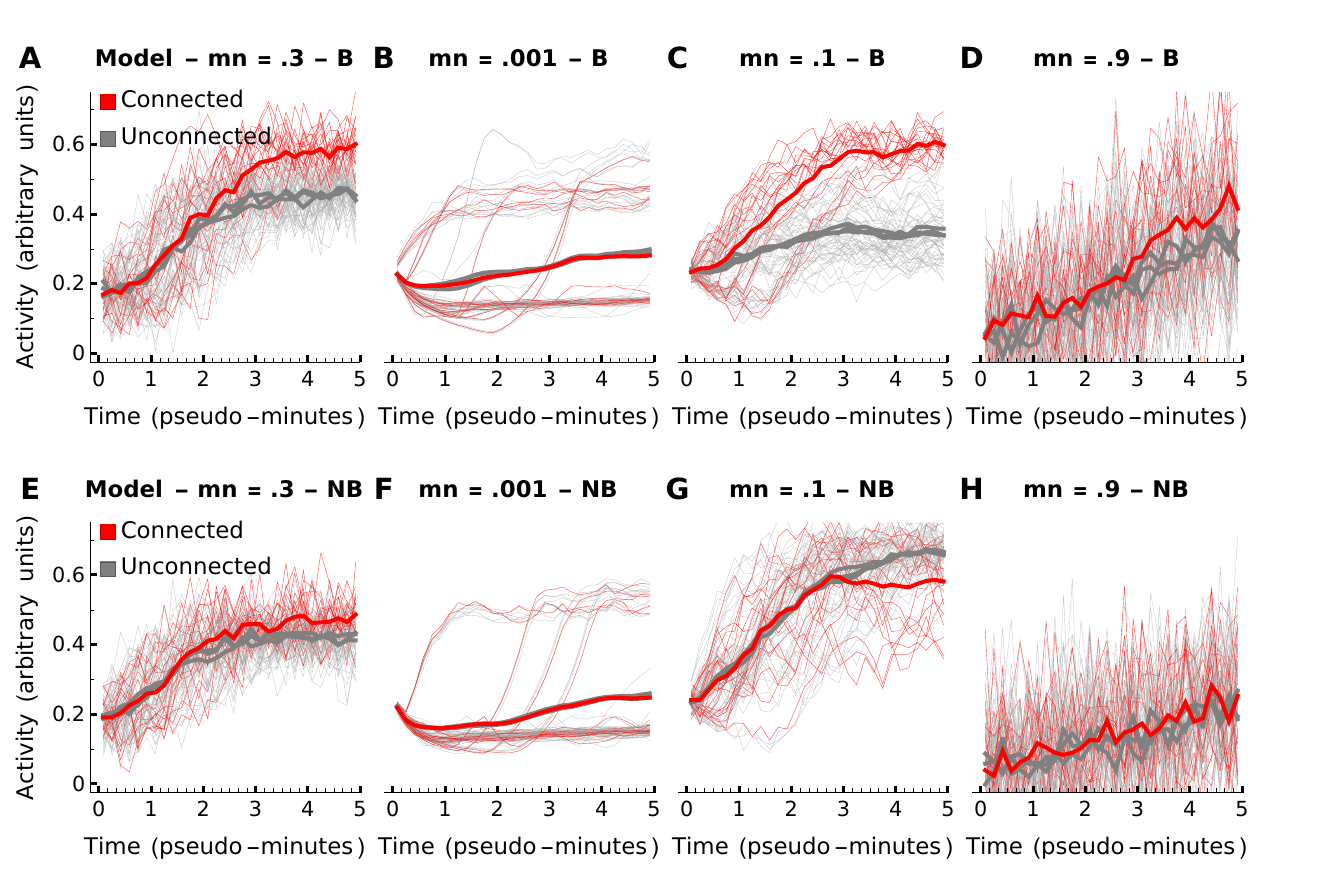}
    \caption{\textbf{Changing the level of motor noise.} The thick curves show the mean activities (mean activity per limb and per 10-s bin, data on 20 simulation runs), across individual data, of the connected limb (red) and the unconnected limbs (gray). Thinner pale curves show individual simulation runs. (A) Complete model in the Binary condition with motor noise mn~=~0.3. (B-D) Ablated models in the Binary condition with motor noise of 0.001, 0.1 and 0.9, respectively.  (E) Complete model in the Non-Binary condition with motor noise mn~=~0.3. (F-H) Ablated models the Non-Binary condition with motor noise of 0.001, 0.1 and 0.9, respectively.}
    \label{fig:MotorAb}
\end{figure}

In this ablation study, shown in Fig. \ref{fig:MotorAb}, we explored the effect of changing the amount of motor noise from the value of 0.3 used in the complete model to very low values of 0.01 and 0.1 and a very high value of 0.9. We observe that the model's learning process is highly sensitive to motor noise. To achieve optimal learning, motor noise is essential; however, both insufficient and excessive levels of motor noise are detrimental.

As shown in Fig. \ref{fig:MotorAb} B and F, the ablation with a motor noise of 0.001 stops the model from learning limb differentiation, with the connected and unconnected limbs behaving very similarly, with a lot of individual variation. There are additionally some individual cases with almost no exploration. 

We suggest the following explanation. As the neural network and beta function are initialized from the same uniform distribution for all limbs, the model will likely start with a very similar activity for each of the four limbs. As a consequence, the losses will change the weights of each neuron similarly, and since the neuron weights evolve from the losses, they will evolve in a similar way. This means that it will be difficult to detect the connected limb. However, if there is significant motor noise, the limbs will not move the same way, meaning they will be affected differently by the loss, allowing the model to eventually figure out which of them is the connected one.

In the ablation in which the motor noise was lowered to 0.1 (Fig. \ref{fig:MotorAb} C and G), we observe a stark difference between the binary condition and the non-binary condition. In the binary condition, the model quickly learns to identify the connected limb and significantly increases the activity in that limb compared to the unconnected limbs. In contrast, in the non-binary condition, the model seems to have more problems identifying the connected limb. For the first 2.5 pseudo-minutes of the non-binary condition, the connected and unconnected limbs have very similar activity levels. After that, surprisingly, the connected limb has lower activity than the unconnected limbs. The overall activity of the unconnected limbs is also higher than without ablation, and the overall variability in activity of all limbs is increased. 

Finally, in the ablation in which the motor noise was raised to 0.9 (Fig. \ref{fig:MotorAb} D and H), learning is hindered even more. The network has very little control over its actions because they become essentially random. This lack of control prevents learning: the average activity of the connected limb and the unconnected limbs progresses similarly, although there is some differentiation of the connected limb in the binary condition. A point worth noting is that when the motor noise is as high as 0.9, the model activity starts lower than 0.2. We assume this occurs because, with such high motor noise, the model will not have attained the correct baseline of about 0.2, even after the initialization period of 1000 steps. 

\begin{figure}
    \centering
    \includegraphics[width=\linewidth]{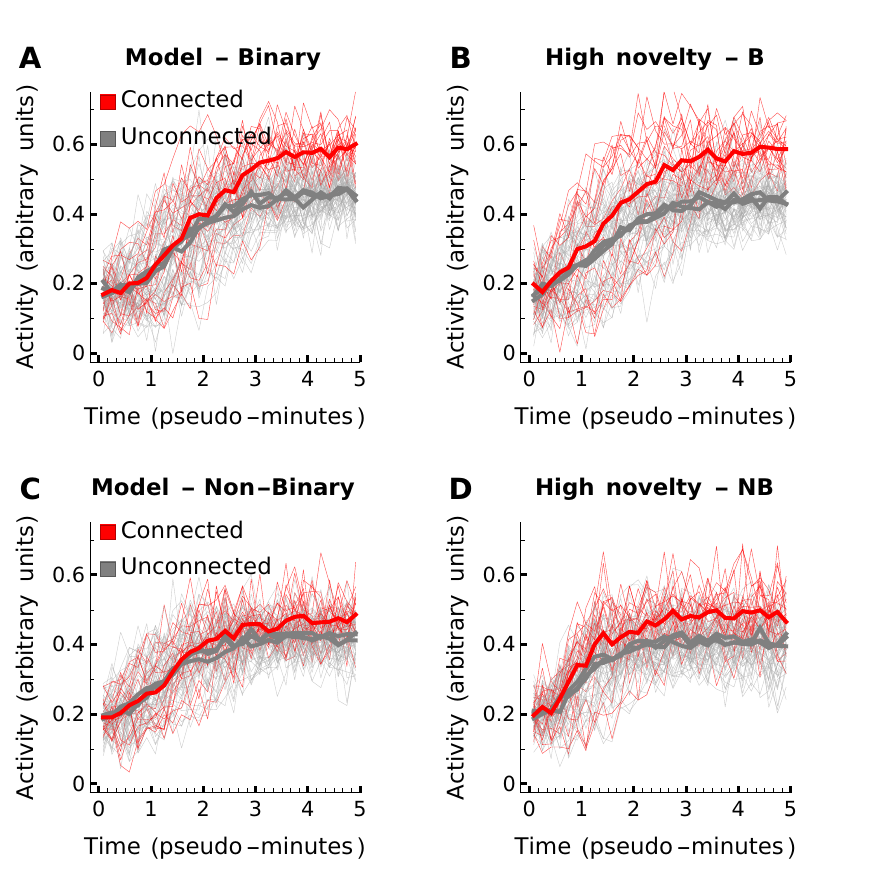}
    \caption{\textbf{Increasing the novelty threshold.} The thick curves show the mean activities (mean activity per limb and per 10-s bin, data on 20 simulation runs), across individual data, of the connected limb (red) and the unconnected limbs (gray). Thinner pale curves show individual simulation runs. (A) Complete model in the Binary condition. (B) Ablated model in the Binary condition. (C) Complete model in the Non-Binary condition. (D) Ablated model the Non-Binary condition.}
    \label{fig:noveltyAb}
\end{figure}

\subsubsection{Increasing the novelty threshold}
In this ablation study we increased the novelty threshold (from 0.1 to 0.3). The novelty threshold determines whether the model interprets a sensory feedback as surprising or not. The main results of this ablation study is that a higher novelty thresholds leads to a somewhat higher variability and an earlier differentiation of the connected and unconnected limbs, as shown in Fig.~\ref{fig:noveltyAb}. In the non-binary condition the final difference in activity of the limbs also seems somewhat higher. Presumably, these effects arise because increasing the novelty threshold decreases the effect of motor noise and makes learning easier. It may be possible that carefully adjusting the novelty threshold could at least partially allow us to match the increased variability observed overall in infants, including in the binary condition and from the very beginning of the experiment.

\begin{figure}
\includegraphics[width=\linewidth]{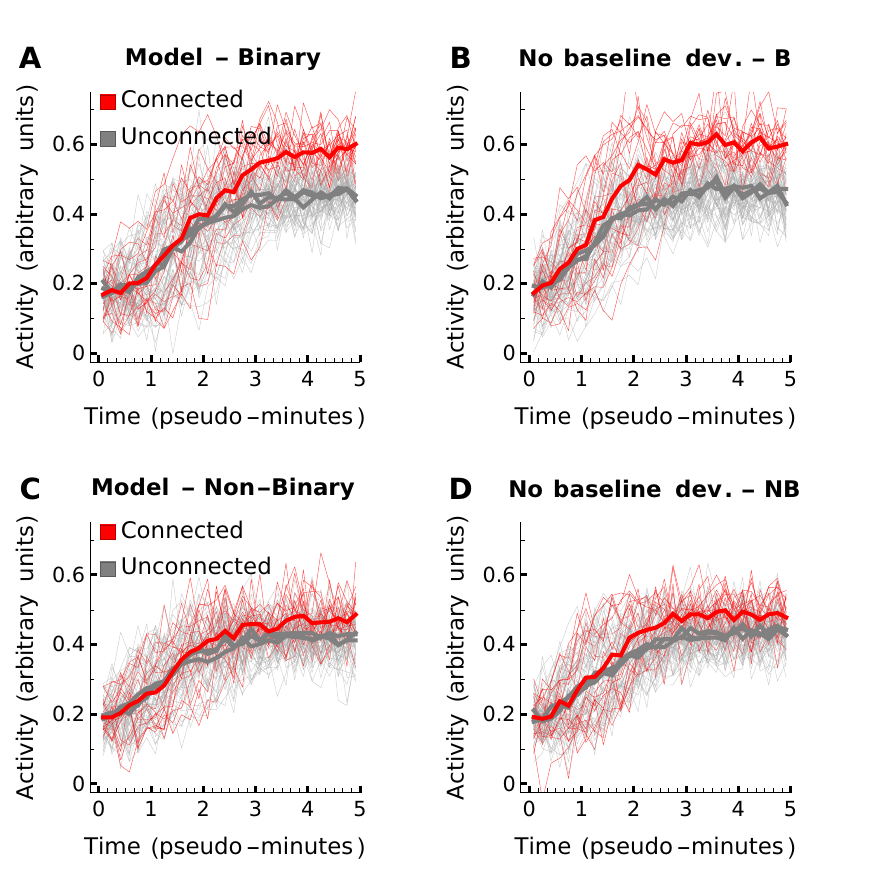}
    \caption{\textbf{Ablation of the deviation from baseline activity loss.} The thick curves show the mean activities (mean activity per limb and per 10-s bin, data on 20 simulation runs), across individual data, of the connected limb (red) and the unconnected limbs (gray). Thinner pale curves show individual simulation runs. (A) Complete model in the Binary condition with the novelty threshold set at 0.1. (B) Ablated model in the Binary condition with the novelty threshold increased to 0.3. (C) Complete model in the Non-Binary condition with the novelty threshold set at 0.1. (D) Ablated model the Non-Binary condition with the novelty threshold increased to 0.3.}
    \label{fig:baselineAb}
\end{figure}

\subsubsection{No deviation from baseline activity error}

In this ablation study (Fig. \ref{fig:baselineAb}), we remove the deviation from baseline activity loss. The condition without the baseline activity loss shows little difference compared to the condition with the baseline activity loss. The reason for introducing the deviation from baseline activity loss in our model was twofold: to limit the overall activity to some activity range around the baseline level and to account for the gradual increase in fussiness observed in infants as the experiments progressed. Presumably, the reason for the small effect of this ablation is that the urge to explore different activity ranges is the main driver of limb activity, and adding a factor that was supposed to simulate ``fussiness'' over time was superfluous.

\section{Discussion, Conclusion, and Future Work}
This paper introduced a computational model that simulates key mechanisms of contingency learning in the “mobile paradigm”. In this paradigm, an infant observes movements or listens to sounds produced by a 'mobile' connected to one of its limbs. It has been observed that infants soon move the \textit{connected} limb more than others. We compare model simulations with the behavior classically observed in the literature and with behavior of 6-month-old infants in two recent studies: one using a \textit{conjugate} (\textit{non-binary}) version where the sensory feedback strength varied with the activity of the connected limb \cite{jacquey_popescu_2020} and another using a \textit{binary} version where the mobile was activated for a fixed time once activity of the connected limb passed a threshold \cite{popescu_6-month-old_2021}. Both studies included a \textit{contingent} group (mobile response tied to limb movement) and a \textit{non-contingent} control group. In addition to replicating these results in our simulations, we also investigated the presence of an \textit{extinction burst}---an increase in activity when the connection is removed---a phenomenon reported in some studies \cite{alessandri_violation_1990},\cite{heathcock_relative_2005}, \cite{lewis_violation_1990},\cite{rovee-collier_topographical_1978} but not consistently observed. 

Our model of the mobile paradigm incorporates components that we believe are important in real infants: an action-outcome prediction mechanism and an exploration mechanism, as well as motor noise and motor control involving multiple muscle commands. However, our current model excludes attentional, social, and environmental factors. 

The results of our simulations show that for the most part our model exhibits behavior similar to that of infants. In the contingent group, the model rapidly differentiates between the connected and unconnected limbs and increases the activity of its connected limb relative to the unconnected limbs. The model also correctly simulates the higher activity observed for infants in a contingent group compared to a non-contingent control group. Our model also simulates the non-systematic emergence of an extinction burst following the disconnection of contingency. Without any modification, the model also correctly simulates the slightly more robust differentiation of the connected versus unconnected limb in the binary variant of the paradigm compared to the non-binary conjugate variant. 

Of course, there are differences between the behavior of the model and infants (see below subsection \ref{limitations_and_future}). Nonetheless, we hypothesize that our model and ablation studies demonstrate that certain mechanisms in our model appear to be essential to simulate infant behavior correctly. This suggests that similar developmental mechanisms could be necessary in infants.

\subsection{Essential components of our model}

Internal \textbf{prediction of sensory effects} was a component suggested by contemporary literature on motor control and intrinsic motivation \cite{wolpert_motor_2001}, \cite{friston_world_2021},\cite{poli_curiosity_2024}. Our ablation studies show that without this component the model cannot learn to differentiate the connected and unconnected limbs. The prediction mechanism strongly affects exploration, and removing it prevents the model from properly evaluating the effectiveness of its exploration.
This exploration itself is supported by a second essential component of our model---the \textbf{novelty-based reinforcement calculator}. Its purpose is to encourage the model to explore new activity levels of its limbs. This mechanism proved essential for the model to learn which of the limbs is connected to the mobile.

To correctly simulate infant behavior, the model also appears to need a large number of \textbf{muscle command neurons}, thereby simulating a population code innervating multiple muscles that control each limb. The ablation studies show that without the numerous muscle command neurons, the model cannot identify the connected limb. Our intuition is that having multiple ways of moving the limb provides more opportunities for the network to discover an effective movement. This is consistent with some results observed in robot learning, where controlling joint angles provided more accurate learning than directly controlling the position of the end effector \cite{spisak2025dirigent}.

\textbf{Motor noise} is another component suggested in the literature (see recent reviews \cite{casartelli_neural_2023} and \cite{gliga_telling_2018}) that diversifies exploration by adding variability. Our ablation study showed that while high motor noise and low motor noise can both inhibit learning, a specific intermediate amount of motor noise is needed to enable quick learning. This finding is consistent with a computational model by Caligiore et al.  \cite{caligiore_integrating_2014} showing that some variability supports exploration and learning. It is also consistent with the finding by Ossmy et al. \cite{ossmy_variety_2018} that in a robotic simulation of soccer, similarly to when infants learn to walk, greater variability leads to better performance.

\subsection{Limitations of our model and future work} \label{limitations_and_future}

An aspect of infant behavior that our simulations did not replicate well is the variability between individual infants as well as within a single infant. Observing less variability across different simulations is expected because our model is designed to imitate an individual infant, whereas behavioral data is obtained from many different infants. However, the lower variability within individual simulation runs seems to be a limitation of our model. As observed in our ablation studies, tripling the value of the novelty threshold increased the variability of individual runs. Thus, adjusting the novelty threshold of the model would be one way to make the model behave more similarly to infants. 

Our lower observed variability must also be understood from the fact that our model is explicitly designed to be focused solely on the task at hand, whereas real infants are subject to multiple distractions during the course of an experiment and are presumably switching their cognitive resources from one task to another. Future work on our model could implement such concurrent learning opportunities and a mechanism that allows task switching. Similar ideas have been advanced, for example, in \cite{oudeyer2007intrinsic} and \cite{butko_learningToLearn_2007}. Note that variability could also be related to the issues of habituation  \cite{sirois_habituation_models_2002}, boredom, fussiness, loss of interest, and, more generally, attentional fluctuations, which were not considered in our current model.

One other limitation of our model is that we only tested one type of sensorimotor contingency. To determine how general our model is and whether it could learn a wider variety of sensorimotor contingencies encountered in early infancy, it should simulate the transfer of learning to new contingencies or with different environmental variables. For example, future experiments could test the introduction of a time delay between the motor action and its sensory consequences, changing which limbs are connected, a change in the specificity of motor actions required (e.g., specific angle of knee flexion as in \cite{angulo-kinzler_three-month-old_2002}), or a change in the schedule of contingency (e.g. from deterministic to probabilistic or random).

In conclusion, our model has established that motor noise, exploration, and motor control involving multiple degrees of freedom are important for sensorimotor learning. The model could be extended to include the role of habituation, boredom and attentional fluctuations so as to replicate observed infant variability. By replicating the main findings of the mobile paradigm, our model provides a strong starting point for further exploration of the mechanisms of sensorimotor learning in infants and applications to robotics.

\bibliographystyle{ieeetr} 
\bibliography{bibtex/bib}

@book{baldassarre2013intrinsically,
  title={Intrinsically motivated learning in natural and artificial systems},
  author={Baldassarre, Gianluca and Mirolli, Marco and others},
  year={2013},
  publisher={Springer}
}

@article{fujihira2023dynamical,
  title={Dynamical systems model of development of the action differentiation in early infancy: a requisite of physical agency},
  author={Fujihira, Ryo and Taga, Gentaro},
  journal={Biological Cybernetics},
  volume={117},
  number={1},
  pages={81--93},
  year={2023},
  publisher={Springer}
}

@article{heathcockPerformanceInfantsBorn2004,
  title = {The Performance of Infants Born Preterm and Full-Term in the Mobile Paradigm: Learning and Memory},
  shorttitle = {The Performance of Infants Born Preterm and Full-Term in the Mobile Paradigm},
  author = {Heathcock, Jill C. and Bhat, Anjana N. and Lobo, Michele A. and Galloway, James C.},
  date = {2004-09},
  journaltitle = {Physical Therapy},
  journal={Physical Therapy},
  shortjournal = {Phys Ther},
  volume = {84},
  number = {9},
  year={2004},
  eprint = {15330694},
  eprinttype = {pmid},
  pages = {808--821},
  issn = {0031-9023},
}

@article{oudeyer2007intrinsic,
  title={What is intrinsic motivation? A typology of computational approaches},
  author={Oudeyer, Pierre-Yves and Kaplan, Frederic},
  journal={Frontiers in neurorobotics},
  volume={1},
  pages={108},
  year={2007},
  publisher={Frontiers}
}

@article{sutton2018reinforcement,
  title={Reinforcement learning: An introduction},
  author={Sutton, Richard S},
  journal={A Bradford Book},
  year={2018}
}

@article{zaadnoordijk_can_2018,
	title = {Can infants' sense of agency be found in their behavior? {Insights} from babybot simulations of the mobile-paradigm},
	volume = {181},
	issn = {0010-0277},
	shorttitle = {Can infants' sense of agency be found in their behavior?},
	url = {http://www.sciencedirect.com/science/article/pii/S0010027718301896},
	doi = {10.1016/j.cognition.2018.07.006},
	language = {en},
	urldate = {2020-09-03},
	journal = {Cognition},
	author = {Zaadnoordijk, Lorijn and Otworowska, Maria and Kwisthout, Johan and Hunnius, Sabine},
	month = dec,
	year = {2018},
	pages = {58--64},
}

@article{zaadnoordijk_movement_2020,
	title = {From movement to action: {An} {EEG} study into the emerging sense of agency in early infancy},
	volume = {42},
	issn = {1878-9293},
	shorttitle = {From movement to action},
	url = {https://www.sciencedirect.com/science/article/pii/S1878929320300086},
	doi = {10.1016/j.dcn.2020.100760},
	urldate = {2024-03-20},
	journal = {Developmental Cognitive Neuroscience},
	author = {Zaadnoordijk, Lorijn and Meyer, Marlene and Zaharieva, Martina and Kemalasari, Falma and van Pelt, Stan and Hunnius, Sabine},
	month = apr,
	year = {2020},
	pages = {100760},
}

@article{watson_reactions_1972,
	title = {Reactions to response-contingent stimulation in early infancy},
	volume = {18},
	issn = {0026-0150},
	url = {https://www.jstor.org/stable/23084608},
	number = {3},
	urldate = {2021-01-06},
	journal = {Merrill-Palmer Quarterly of Behavior and Development},
	author = {Watson, John S. and Ramey, Craig T.},
	year = {1972},
	pages = {219--227},
}

@article{sloan_meaning_2023,
	title = {Meaning from movement and stillness: {Signatures} of coordination dynamics reveal infant agency},
	volume = {120},
	shorttitle = {Meaning from movement and stillness},
	url = {https://www.pnas.org/doi/10.1073/pnas.2306732120},
	doi = {10.1073/pnas.2306732120},
	number = {39},
	urldate = {2023-10-04},
	journal = {Proceedings of the National Academy of Sciences},
	author = {Sloan, Aliza T. and Jones, Nancy Aaron and Kelso, J. A. Scott},
	month = sep,
	year = {2023},
	pages = {e2306732120},
}

@article{sen_methodological_2024,
	title = {Methodological integrity assessment in the mobile paradigm literature: {A} lesson for understanding opportunistic use of researcher degrees of freedom in psychology},
	volume = {95},
	issn = {1467-8624},
	shorttitle = {Methodological integrity assessment in the mobile paradigm literature},
	doi = {10.1111/cdev.13850},
	number = {2},
	journal = {Child Development},
	author = {Sen, Umay and Gredebäck, Gustaf},
	year = {2024},
	pages = {338--353},
}

@article{rovee_conjugate_1969,
	title = {Conjugate reinforcement of infant exploratory behavior},
	volume = {8},
	url = {http://www.sciencedirect.com/science/article/pii/0022096569900253},
	doi = {https://doi.org/10.1016/0022-0965(69)90025-3},
	number = {1},
	urldate = {2016-06-02},
	journal = {Journal of experimental child psychology},
	author = {Rovee, Carolyn Kent and Rovee, David T.},
	year = {1969},
	note = {00334
Citation Key Alias: rovee1969},
	pages = {33--39},
}

@article{sen_making_2021,
	title = {Making the {World} {Behave}: {A} {New} {Embodied} {Account} on {Mobile} {Paradigm}},
	volume = {15},
	issn = {1662-5137},
	shorttitle = {Making the {World} {Behave}},
	url = {https://www.frontiersin.org/articles/10.3389/fnsys.2021.643526},
	doi = {10.3389/fnsys.2021.643526},
	language = {English},
	urldate = {2024-03-26},
	journal = {Frontiers in Systems Neuroscience},
	author = {Sen, Umay and Gredebäck, Gustaf},
	month = mar,
	year = {2021},
	note = {Publisher: Frontiers},
}

@article{rovee-collier_topographical_1978,
	title = {Topographical response differentiation and reversal in 3-month-old infants},
	volume = {1},
	issn = {0163-6383},
	url = {https://www.sciencedirect.com/science/article/pii/S0163638378800447},
	doi = {10.1016/S0163-6383(78)80044-7},
	urldate = {2024-03-26},
	journal = {Infant Behavior and Development},
	author = {Rovee-Collier, Carolyn Kent and Morrongiello, Barbara A. and Aron, Mark and Kupersmidt, Janis},
	month = jan,
	year = {1978},
	pages = {323--333},
}

@article{popescu_6-month-old_2021,
	title = {6-{Month}-{Old} {Infants}’ {Sensitivity} to {Contingency} in a {Variant} of the {Mobile} {Paradigm} {With} {Proximal} {Stimulation} {Studied} at {Fine} {Temporal} {Resolution} in the {Laboratory}},
	volume = {12},
	issn = {1664-1078},
	url = {https://www.frontiersin.org/articles/10.3389/fpsyg.2021.610002/full},
	doi = {10.3389/fpsyg.2021.610002},
	language = {English},
	urldate = {2021-03-30},
	journal = {Frontiers in Psychology},
	author = {Popescu, Sergiu Tcaci and Dauphin, Alice and Vergne, Judith and O’Regan, J. Kevin},
	year = {2021},
	note = {Publisher: Frontiers},
}

@article{poli_curiosity_2024,
	title = {Curiosity and the dynamics of optimal exploration},
	volume = {0},
	issn = {1364-6613, 1879-307X},
	url = {https://www.cell.com/trends/cognitive-sciences/abstract/S1364-6613(24)00028-7},
	doi = {10.1016/j.tics.2024.02.001},
	language = {English},
	number = {0},
	urldate = {2024-03-13},
	journal = {Trends in Cognitive Sciences},
	author = {Poli, Francesco and O’Reilly, Jill X. and Mars, Rogier B. and Hunnius, Sabine},
	month = feb,
	year = {2024},
	pmid = {38413257},
	note = {Publisher: Elsevier},
}

@article{ossmy_variety_2018,
	title = {Variety {Wins}: {Soccer}-{Playing} {Robots} and {Infant} {Walking}},
	volume = {12},
	issn = {1662-5218},
	shorttitle = {Variety {Wins}},
	url = {https://www.frontiersin.org/article/10.3389/fnbot.2018.00019},
	urldate = {2022-01-14},
	journal = {Frontiers in Neurorobotics},
	author = {Ossmy, Ori and Hoch, Justine E. and MacAlpine, Patrick and Hasan, Shohan and Stone, Peter and Adolph, Karen E.},
	year = {2018},
}

@article{lewis_violation_1990,
	title = {Violation of expectancy, loss of control, and anger expressions in young infants.},
	volume = {26},
	url = {http://psycnet.apa.org/journals/dev/26/5/745/},
	number = {5},
	urldate = {2017-01-23},
	journal = {Developmental Psychology},
	author = {Lewis, Michael and Alessandri, Steven M. and Sullivan, Margaret W.},
	year = {1990},
	pages = {745},
}

@article{jacquey_popescu_2020,
	title = {Development of body knowledge as measured by arm differentiation in infants: {From} global to local?},
	volume = {38},
	copyright = {© 2019 The Authors. British Journal of Developmental Psychology published byJohn Wiley\&Sons Ltd on behalf of British Psychological Society},
	issn = {2044-835X},
	shorttitle = {Development of body knowledge as measured by arm differentiation in infants},
	url = {http://onlinelibrary.wiley.com/doi/abs/10.1111/bjdp.12309},
	doi = {10.1111/bjdp.12309},
	language = {en},
	number = {1},
	urldate = {2020-07-04},
	journal = {British Journal of Developmental Psychology},
	author = {Jacquey, Lisa and Popescu, Sergiu Tcaci and Vergne, Judith and Fagard, Jacqueline and Esseily, Rana and O’Regan, Kevin},
	year = {2020},
	note = {\_eprint: https://onlinelibrary.wiley.com/doi/pdf/10.1111/bjdp.12309},
	pages = {108--124},
}

@article{kelso_coordination_2016,
	title = {The coordination dynamics of mobile conjugate reinforcement},
	volume = {110},
	issn = {1432-0770},
	url = {https://doi.org/10.1007/s00422-015-0676-0},
	doi = {10.1007/s00422-015-0676-0},
	language = {en},
	number = {1},
	urldate = {2023-05-19},
	journal = {Biological Cybernetics},
	author = {Kelso, J. A. Scott and Fuchs, Armin},
	month = feb,
	year = {2016},
	pages = {41--53},
}

@article{jacquey_fagard_2020,
	title = {Detection of sensorimotor contingencies in infants before the age of 1 year: {A} comprehensive review},
	volume = {56},
	issn = {1939-0599},
	shorttitle = {Detection of sensorimotor contingencies in infants before the age of 1 year},
	doi = {10.1037/dev0000916},
	language = {eng},
	number = {7},
	journal = {Developmental Psychology},
	author = {Jacquey, Lisa and Fagard, Jacqueline and Esseily, Rana and O'Regan, J. Kevin},
	month = jul,
	year = {2020},
	pmid = {32463268},
	pages = {1233--1251},
}

@article{heathcock_relative_2005,
	title = {The {Relative} {Kicking} {Frequency} of {Infants} {Born} {Full}-term and {Preterm} {During} {Learning} and {Short}-term and {Long}-term {Memory} {Periods} of the {Mobile} {Paradigm}},
	volume = {85},
	journal = {Physical Therapy},
	author = {Heathcock, Jill C. and Bhat, Anjana N. and Lobo, Michele A. and Galloway, James C},
	year = {2005},
	pages = {8--18},
}

@article{gliga_telling_2018,
	title = {Telling {Apart} {Motor} {Noise} and {Exploratory} {Behavior}, in {Early} {Development}},
	volume = {9},
	issn = {1664-1078},
	url = {https://www.frontiersin.org/article/10.3389/fpsyg.2018.01939},
	urldate = {2022-01-13},
	journal = {Frontiers in Psychology},
	author = {Gliga, Teodora},
	year = {2018},
}

@article{friston_world_2021,
	title = {World model learning and inference},
	issn = {0893-6080},
	url = {https://www.sciencedirect.com/science/article/pii/S0893608021003610},
	doi = {10.1016/j.neunet.2021.09.011},
	language = {en},
	urldate = {2021-09-27},
	journal = {Neural Networks},
	author = {Friston, Karl and Moran, Rosalyn J. and Nagai, Yukie and Taniguchi, Tadahiro and Gomi, Hiroaki and Tenenbaum, Josh},
	month = sep,
	year = {2021},
}

@book{edwards_motor_2010,
	address = {Belmont, CA},
	edition = {1st edition},
	title = {Motor {Learning} and {Control}: {From} {Theory} to {Practice}},
	isbn = {978-0-495-01080-7},
	shorttitle = {Motor {Learning} and {Control}},
	language = {English},
	publisher = {Cengage Learning},
	author = {Edwards, William H.},
	month = aug,
	year = {2010},
}

@article{casartelli_neural_2023,
	title = {From neural noise to co-adaptability: {Rethinking} the multifaceted architecture of motor variability},
	volume = {47},
	issn = {1571-0645},
	shorttitle = {From neural noise to co-adaptability},
	url = {https://www.sciencedirect.com/science/article/pii/S1571064523001793},
	doi = {10.1016/j.plrev.2023.10.036},
	urldate = {2024-06-03},
	journal = {Physics of Life Reviews},
	author = {Casartelli, Luca and Maronati, Camilla and Cavallo, Andrea},
	month = dec,
	year = {2023},
	pages = {245--263},
}

@article{caligiore_integrating_2014,
	title = {Integrating reinforcement learning, equilibrium points, and minimum variance to understand the development of reaching: a computational model},
	volume = {121},
	issn = {1939-1471},
	shorttitle = {Integrating reinforcement learning, equilibrium points, and minimum variance to understand the development of reaching},
	doi = {10.1037/a0037016},
	language = {eng},
	number = {3},
	journal = {Psychological Review},
	author = {Caligiore, Daniele and Parisi, Domenico and Baldassarre, Gianluca},
	month = jul,
	year = {2014},
	pmid = {25090425},
	pages = {389--421},
}

@article{butko_detecting_2010,
	title = {Detecting contingencies: an infomax approach},
	volume = {23},
	issn = {1879-2782},
	shorttitle = {Detecting contingencies},
	doi = {10.1016/j.neunet.2010.09.001},
	language = {eng},
	number = {8-9},
	journal = {Neural Networks: The Official Journal of the International Neural Network Society},
	author = {Butko, Nicholas J. and Movellan, Javier R.},
	year = {2010},
	pmid = {20951334},
	pages = {973--984},
}

@article{bednarski_infants_2022,
	title = {Do infants have agency? {The} importance of control for the study of early agency},
	volume = {64},
	issn = {0273-2297},
	shorttitle = {Do infants have agency?},
	url = {https://www.sciencedirect.com/science/article/pii/S0273229722000120},
	doi = {10.1016/j.dr.2022.101022},
	language = {en},
	urldate = {2023-05-19},
	journal = {Developmental Review},
	author = {Bednarski, Florian Markus and Musholt, Kristina and Grosse Wiesmann, Charlotte},
	month = jun,
	year = {2022},
	pages = {101022},
}

@article{alessandri_violation_1990,
	title = {Violation of expectancy and frustration in early infancy},
	volume = {26},
	issn = {1939-0599},
	doi = {10.1037/0012-1649.26.5.738},
	number = {5},
	journal = {Developmental Psychology},
	author = {Alessandri, Steven M. and Sullivan, Margaret W. and Lewis, Michael},
	year = {1990},
	note = {Place: US
Publisher: American Psychological Association},
	pages = {738--744},
}

@article{watanabe_general_2006,
	title = {General to specific development of movement patterns and memory for contingency between actions and events in young infants},
	volume = {29},
	issn = {01636383},
	url = {http://linkinghub.elsevier.com/retrieve/pii/S0163638306000269},
	doi = {10.1016/j.infbeh.2006.02.001},
	language = {en},
	number = {3},
	urldate = {2016-01-13},
	journal = {Infant Behavior and Development},
	author = {Watanabe, Hama and Taga, Gentaro},
	year = {2006},
	pages = {402--422},
}

@article{adamson_stillface_2003,
Author = {Adamson, LB and Frick, JE},
Title = {The still face: A history of a shared experimental paradigm},
Journal = {INFANCY},
Year = {2003},
Volume = {4},
Number = {4},
Pages = {451-473},
Note = {International Conference on Infant Studies, TORONTO, CANADA, APR, 2002},
Publisher = {WILEY},
Address = {111 RIVER ST, HOBOKEN 07030-5774, NJ USA},
Type = {Article; Proceedings Paper},
Language = {English},
Affiliation = {Georgia State Univ, Dept Psychol, Univ Plaza, Atlanta, GA 30303 USA.
   Georgia State Univ, Dept Psychol, Atlanta, GA 30303 USA.
   Univ Georgia, Dept Psychol, Athens, GA 30602 USA.},
DOI = {10.1207/S15327078IN0404\_01},
ISSN = {1525-0008},
EISSN = {1532-7078},
Keywords-Plus = {INFANT AFFECT; TACTILE STIMULATION; AFFECTIVE RESPONSES; SOCIAL
   COMPETENCE; YOUNG INFANTS; MOTHERS; 3-MONTH-OLD; BEHAVIOR; ATTENTION;
   DEAF},
Research-Areas = {Psychology},
Web-of-Science-Categories  = {Psychology, Developmental},
Author-Email = {ladamson@gsu.edu},
Affiliations = {University System of Georgia; Georgia State University; University
   System of Georgia; University of Georgia},
Number-of-Cited-References = {98},
Times-Cited = {263},
Usage-Count-Last-180-days = {3},
Usage-Count-Since-2013 = {39},
Journal-ISO = {Infancy},
Doc-Delivery-Number = {757VF},
Web-of-Science-Index = {Social Science Citation Index (SSCI); Conference Proceedings Citation Index - Social Science &amp; Humanities (CPCI-SSH)},
Unique-ID = {WOS:000187581600001},
DA = {2025-03-09},
}

@article{tarabulsy_contingency_1996,
	title = {Contingency detection and the contingent organization of behavior in interactions: {Implications} for socioemotional development in infancy},
	volume = {120},
	issn = {1939-1455},
	shorttitle = {Contingency detection and the contingent organization of behavior in interactions},
	doi = {10.1037/0033-2909.120.1.25},
	number = {1},
	journal = {Psychological Bulletin},
	author = {Tarabulsy, George M. and Tessier, Réjean and Kappas, Arvid},
	year = {1996},
	pages = {25--41},
}

@article{greco_ontogeny_1986,
	title = {Ontogeny of {Early} {Event} {Memory}: {I}. {Forgetting} and {Retrieval} by 2- and 3-{Month}-{Olds}},
	volume = {9},
	journal = {Infant Behavior and Development},
	author = {Greco, Carolyn and Rovee-Collier, Carolyn K. and Hayne, Harlene and Griesler, Pamela C. and Earley, Linda A.},
	year = {1986},
	pages = {441--460},
}

@article{hayne_ontogeny_1986,
	title = {Ontogeny of {Early} {Event} {Memory}: {II}. {Encoding} and {Retrieval} by 2- and 3-{Month}-{Olds}},
	volume = {9},
	journal = {Infant Behavior and Development},
	author = {Hayne, Harlene and Greco, Carolyn and Earley, Linda A. and Griesler, Pamela C. and Rovee-Collier, Carolyn},
	year = {1986},
	pages = {461--472},
}

@article{lewis_emotional_1985,
	title = {Emotional behaviour during the learning of a contingency in early infancy},
	volume = {3},
	issn = {0261510X},
	url = {http://doi.wiley.com/10.1111/j.2044-835X.1985.tb00982.x},
	doi = {10.1111/j.2044-835X.1985.tb00982.x},
	language = {en},
	number = {3},
	urldate = {2019-04-16},
	journal = {British Journal of Developmental Psychology},
	author = {Lewis, Michael and Sullivan, Margaret Wolan and Brooks-Gunn, Jeanne},
	month = sep,
	year = {1985},
	pages = {307--316},
}

@article{rovee-collier_development_1999,
	title = {The {Development} of {Infant} {Memory}},
	volume = {8},
	language = {en},
	number = {3},
	journal = {Current Directions in Psychological Science},
	author = {Rovee-Collier, Carolyn},
	year = {1999},
	pages = {6},
}

@article{rovee-collier_long-term_1999,
	title = {Long-term maintenance of infant memory},
	volume = {35},
	issn = {0012-1630, 1098-2302},
	url = {http://doi.wiley.com/10.1002/%28SICI%291098-2302%28199909%2935%3A2%3C91%3A%3AAID-DEV2%3E3.0.CO%3B2-U},
	doi = {10.1002/(SICI)1098-2302(199909)35:2<91::AID-DEV2>3.0.CO;2-U},
	language = {en},
	number = {2},
	urldate = {2019-05-28},
	journal = {Developmental Psychobiology},
	author = {Rovee-Collier, Carolyn and Hartshorn, Kristin and DiRubbo, Manda},
	month = sep,
	year = {1999},
	pages = {91--102},
}

@inproceedings{movellan_Watson_development_2002,
	title = {The development of gaze following as a {Bayesian} systems identification problem},
	url = {http://ieeexplore.ieee.org/xpls/abs_all.jsp?arnumber=1011728},
	urldate = {2016-01-26},
	booktitle = {Development and {Learning}, 2002. {Proceedings}. {The} 2nd {International} {Conference} on},
	publisher = {IEEE},
	author = {Movellan, Javier R. and Watson, John S.},
	year = {2002},
	pages = {34--40},
}

@inproceedings{butko_learningToLearn_2007,
	title = {Learning to learn},
	doi = {10.1109/DEVLRN.2007.4354070},
	booktitle = {2007 {IEEE} 6th {International} {Conference} on {Development} and {Learning}},
	author = {Butko, N. J. and Movellan, J. R.},
	month = jul,
	year = {2007},
	pages = {151--156},
}

@article{wolpert_motor_2001,
	title = {Motor prediction},
	volume = {11},
	issn = {0960-9822},
	url = {https://www.sciencedirect.com/science/article/pii/S0960982201004328},
	doi = {10.1016/S0960-9822(01)00432-8},
	number = {18},
	urldate = {2025-04-01},
	journal = {Current Biology},
	author = {Wolpert, Daniel M and Flanagan, J. Randall},
	month = sep,
	year = {2001},
	pages = {R729--R732},
}

@article{sirois_habituation_models_2002,
	title = {Models of habituation in infancy},
	volume = {6},
	issn = {1364-6613, 1879-307X},
	url = {https://www.cell.com/trends/cognitive-sciences/abstract/S1364-6613(02)01926-5},
	doi = {10.1016/S1364-6613(02)01926-5},
	language = {English},
	number = {7},
	urldate = {2025-04-02},
	journal = {Trends in Cognitive Sciences},
	author = {Sirois, Sylvain and Mareschal, Denis},
	month = jul,
	year = {2002},
	pmid = {12110362},
	pages = {293--298},
}

@article{angulo-kinzler_three-month-old_2002,
	title = {Three-{Month}-{Old} {Infants} {Can} {Select} {Specific} {Leg} {Motor} {Solutions}},
	volume = {6},
	issn = {1087-1640, 1543-2696},
	url = {https://journals.humankinetics.com/view/journals/mcj/6/1/article-p52.xml},
	doi = {10.1123/mcj.6.1.52},
	language = {en\_US},
	number = {1},
	urldate = {2020-10-30},
	journal = {Motor Control},
	author = {Angulo-Kinzler, Rosa M. and Ulrich, Beverly and Thelen, Esther},
	month = jan,
	year = {2002},
	pages = {52--68},
}

@article{spisak2025dirigent,
  title={DIRIGENt: End-To-End Robotic Imitation of Human Demonstrations Based on a Diffusion Model},
  author={Spisak, Josua and Kerzel, Matthias and Wermter, Stefan},
  journal={arXiv preprint arXiv:2501.16800},
  year={2025}
}

@article{goddu2024development,
  title={The development of human causal learning and reasoning},
  author={Goddu, Mariel K and Gopnik, Alison},
  journal={Nature Reviews Psychology},
  volume={3},
  number={5},
  pages={319--339},
  year={2024},
  publisher={Nature Publishing Group US New York}
}
\vspace{-1.2cm}
\begin{IEEEbiography}    [{\includegraphics[width=1in,clip,keepaspectratio]{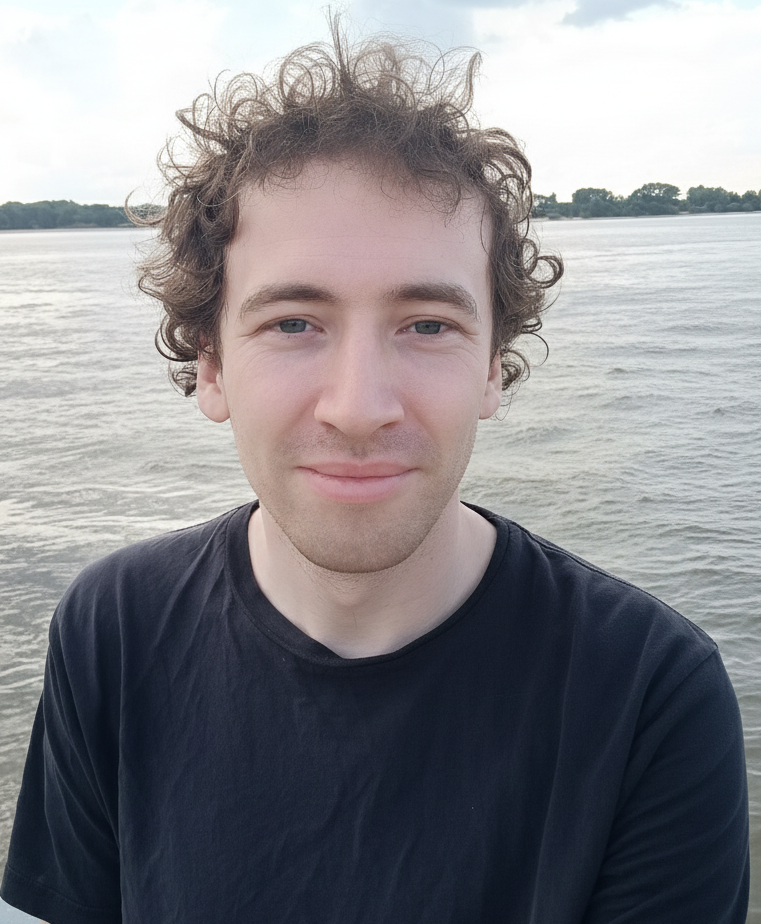}}]{Josua Spisak}
is a Ph.D student at the University of Hamburg, Germany in the Knowledge Technology Institute of the Dept. of Informatics. He is part of the DFG SPP ``The Active Self'' under the MoReSpace project. His research interests lie in the field of machine learning with a particular focus on imitation learning and computational modeling.
\end{IEEEbiography}
\vspace{-1.5cm}
\begin{IEEEbiography}    [{\includegraphics[width=1in,clip,keepaspectratio]{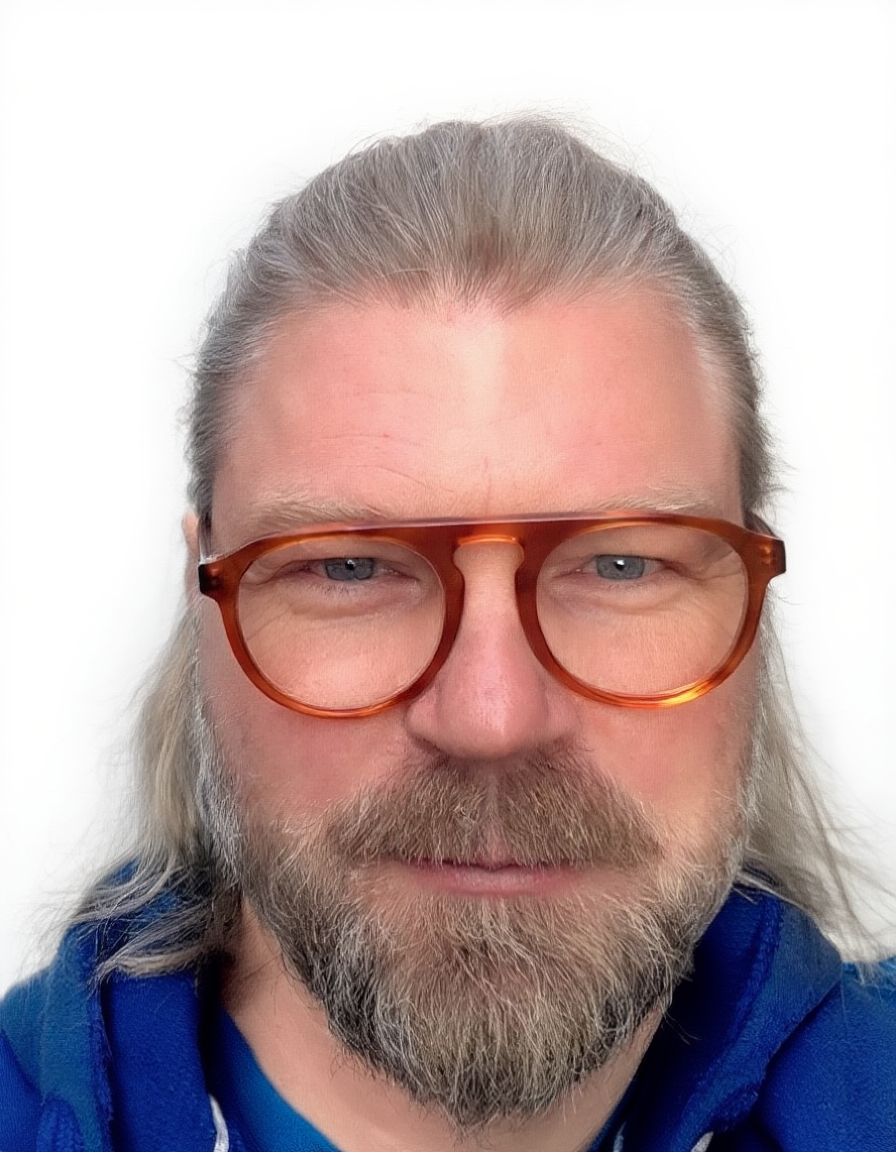}}]{Sergiu Tcaci Popescu} received the Ph.D. degree in Cognitive Science from EHESS, Paris, in 2018. From 2022 to 2024, he was a postdoctoral researcher supported by the Johannes Amos Comenius Research Fellowship, conducting research in developmental psychology and cognitive robotics. He is a Postdoctoral Researcher with the Humanoid and Cognitive Robotics Group at the Czech Technical University in Prague and an Associated Researcher with the Integrative Neuroscience \& Cognition Center (CNRS/Université Paris Cité). His research interests include sensorimotor development in infants, embodied and spatial cognition, contingency and causal learning, and computational and robotic models of infant behavior, with a particular focus on collaborations between cognitive science, computer science, and robotics to study human learning.
\end{IEEEbiography}
\vspace{-1.5cm}
\begin{IEEEbiography}    [{\includegraphics[width=1in,clip,keepaspectratio]{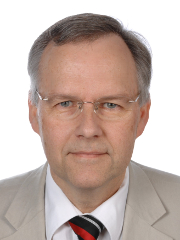}}]{Stefan Wermter}
 is Full Professor at the University of Hamburg, Germany, and Director of the Knowledge Technology Institute in the Dept. of Informatics. He has previously held positions at the Technical University of Dortmund,  University of Massachusetts, the International Computer Science Institute in Berkeley and the University of Sunderland. His main research interests are in neural networks, hybrid knowledge technology,  neuroscience-inspired computing, cognitive robotics, natural language processing and human-robot interaction. He is coordinator of the international doctoral training network TRAIL and he currently serves as the President of the European Neural Network Society.  
\end{IEEEbiography}
\vspace{-1.5cm}
\begin{IEEEbiography}    [{\includegraphics[width=1in,clip,keepaspectratio]{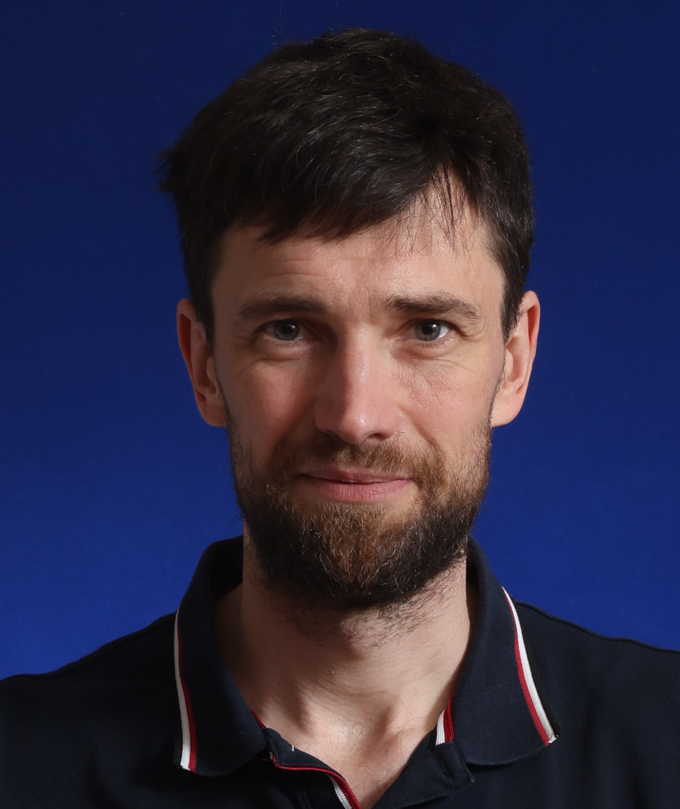}}]{Matej Hoffmann}
(Senior Member, IEEE) received the Ph.D. degree in Informatics from Artificial Intelligence Lab, University of Zurich, Zurich, Switzerland, in 2012. From 2013 to 2016, he conducted postdoctoral research with the iCub Facility of the Italian Institute of Technology, Genoa, Italy, supported by a Marie Curie Intra-European Fellowship. In 2017, he joined the Department of Cybernetics, Faculty of Electrical Engineering, Czech Technical University in Prague, where he is currently an Associate Professor and the Coordinator of the Humanoid and Cognitive Robotics Group. His research interests include humanoids, cognitive robotics, and sensorimotor development in infants.
\end{IEEEbiography}
\vspace{-1.5cm}
\begin{IEEEbiography}    [{\includegraphics[width=1in,clip,keepaspectratio]{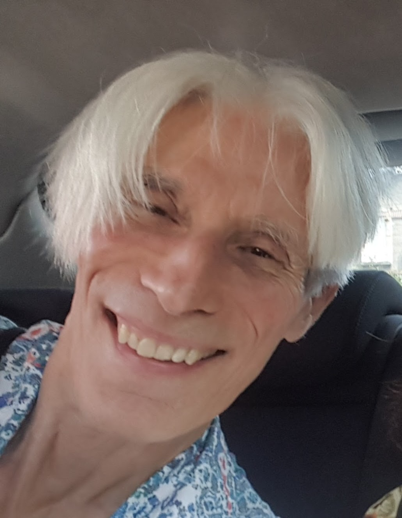}}]{Kevin O'Regan} is emeritus ex-director of the Laboratoire Psychologie de la Perception, CNRS, Université Paris Descartes. After working on eye movements in reading he became interested in visual stability and discovered the phenomenon of change blindness. His current work concerns the sensorimotor approach to phenomenal consciousness and its applications to child development and robotics. See \url{http://whatfeelingislike.net} and \url{http://kevin-oregan.net}.
\end{IEEEbiography}

\clearpage
\section{Appendix}\label{Appendix}

\FloatBarrier
\vspace{0.7cm} 

\begin{figure}[H]
  \vspace{-4.5em}
   \includegraphics[width=\linewidth]{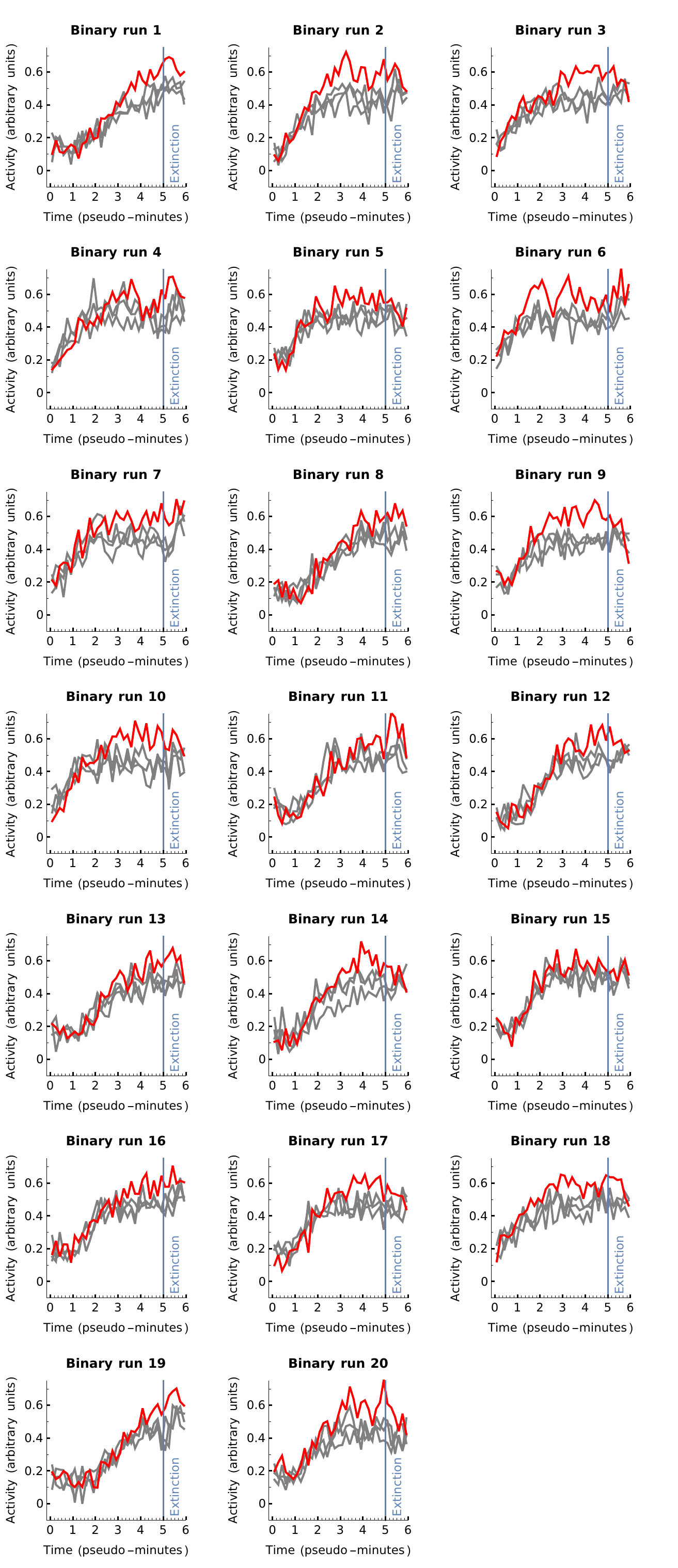}
  \caption{\textbf{Binary model - 20 individual runs.} The activity is shown per limb and per 10-s bin. The red curve shows the activity of the connected limb and the gray curves show the activities of the unconnected limbs. The time of contingency removal is emphasized with a vertical line. 
  The one-minute extinction period after contingency removal is shown on contrasting background.
    }
   \label{fig:individualStandardB}
\end{figure}

\begin{figure}
    \vspace{3em} 
    \includegraphics[width=\linewidth]{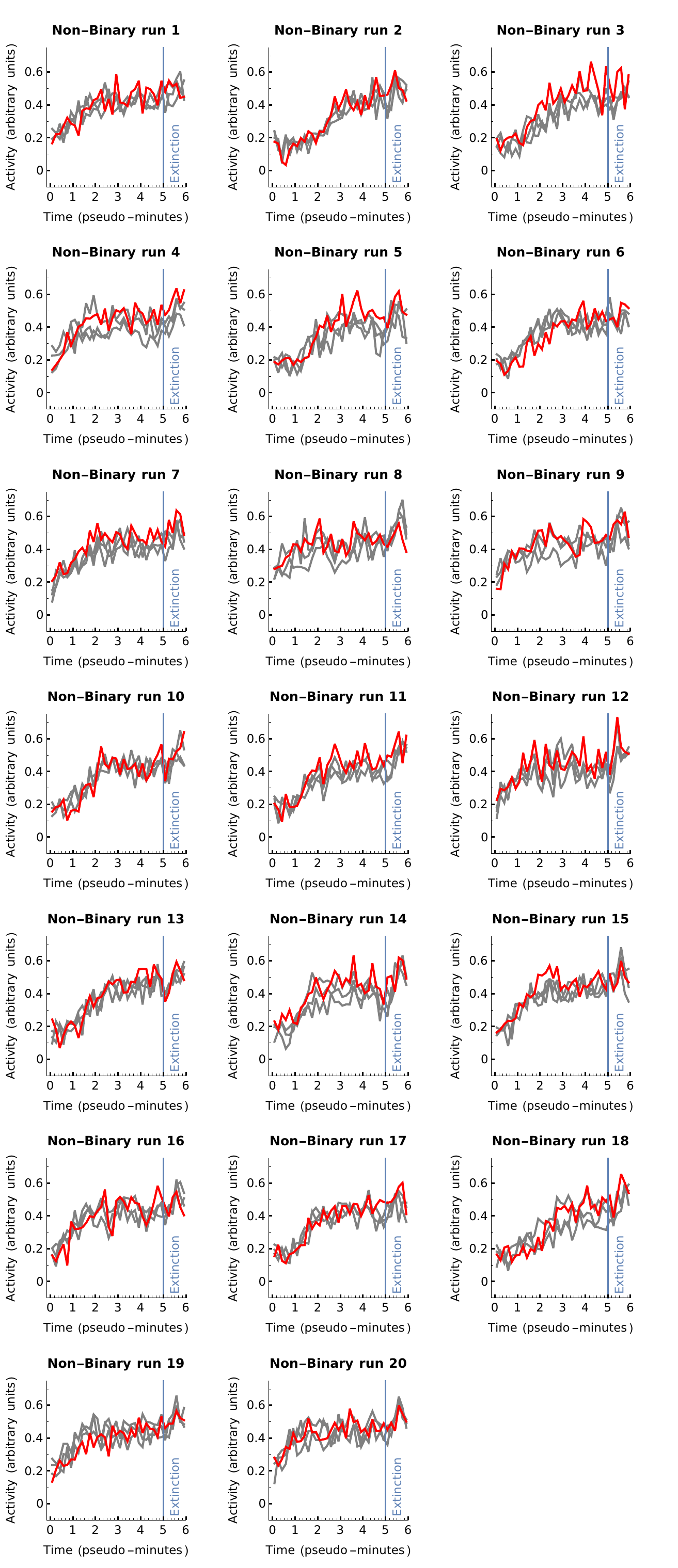}
    \caption{\textbf{Non-Binary model - 20 individual runs.} The activity is shown per limb and per 10-s bin. The red curve shows the activity of the connected limb and the gray curves show the activities of the unconnected limbs. The time of contingency removal is emphasized with a vertical line.}
    \vspace{128cm} 
    \label{fig:individualStandardNB}
\end{figure}

\end{document}